%% file: paper.tex
\documentclass[letterpaper,11pt]{article}

\usepackage{amsmath,amsthm,amssymb,amsfonts}

\usepackage[mono=false]{libertine}
\usepackage{libertinust1math}
\usepackage[T1]{fontenc}

\usepackage[margin=1in]{geometry}
\usepackage{subfiles}
\usepackage{hyperref}
\hypersetup{colorlinks=true, citecolor=blue, linkcolor=red, urlcolor=blue}
\usepackage{enumerate}
\usepackage[shortlabels]{enumitem} 
\usepackage{graphicx}
\graphicspath{{figures/}}
\usepackage{subcaption}
\usepackage{csquotes}
\usepackage{titling}

\usepackage{appendix}

\newtheorem{thm}{Theorem}[section]
\newtheorem*{thm*}{Theorem}
\newtheorem{cor}[thm]{Corollary}
\newtheorem{lem}[thm]{Lemma}
\newtheorem*{lem*}{Lemma}

\theoremstyle{definition}

\newtheorem*{defn*}{Definition}
\theoremstyle{remark}
\newtheorem{rem}{Remark}[section]

\newcommand{\figref}[1]{Figure~\ref{#1}}
\newcommand{\secref}[1]{Section~\ref{#1}}
\newcommand{\thmref}[1]{Theorem~\ref{#1}}
\newcommand{\lemref}[1]{Lemma~\ref{#1}}

\newcommand{\corref}[1]{Corollary~\ref{#1}}

\newcommand{\Holant}{\operatorname{Holant}}
\newcommand{\sub}[1]{\raisebox{-.1\baselineskip}{\includegraphics[height=0.7\baselineskip, width=0.7\baselineskip, keepaspectratio]{sub_#1}}}
\newcommand{\subinmatrix}[1]{\raisebox{-.1\baselineskip}{\includegraphics[height=\baselineskip, width=\baselineskip, keepaspectratio]{sub_#1}}}

\date{}

\begin{document}

\title{Approximability of the Six-vertex Model}
\author{
{Jin-Yi Cai}\thanks{Department of Computer Sciences, University of Wisconsin-Madison. Supported by NSF CCF-1714275.}
\\ \texttt{jyc@cs.wisc.edu}
\and {Tianyu Liu}\thanksmark{1}\\
\texttt{tl@cs.wisc.edu}
\and
{Pinyan Lu}\thanks{Institute for Theoretical Computer Science, School of Information Management and Engineering, Shanghai University of Finance and Economics.}
\\ \texttt{lu.pinyan@mail.shufe.edu.cn}
}

\maketitle

\begin{abstract}
In this paper we take the first step toward a classification of the approximation complexity of the six-vertex model, an object of extensive research in statistical physics.
Our complexity results conform to the phase transition phenomenon from physics.
We show that the approximation complexity of the six-vertex model
behaves dramatically differently on
 the two sides separated by the phase transition threshold.
Furthermore, we present structural properties of the six-vertex model 
on planar graphs for parameter settings that have
known relations to the Tutte polynomial $T(G; x, y)$.
\end{abstract}


\section{Introduction}\label{sec:intro}
\subfile{intro}

\section{Preliminaries}\label{sec:prelim}
\subfile{prelim}

\section{Confinement Theorems}\label{sec:theorems}
\subfile{theorems}

\section{FPRAS}\label{sec:fpras}
\subfile{fpras}

\section{Hardness}\label{sec:hardness}
\subfile{hardness}

\section{Open problems}
\subfile{open}

\bibliography{reference}{}
\bibliographystyle{alpha}

\appendix
\section*{Appendix}\label{sec:appendix}
\subfile{appendix}

\end{document}

%% file: intro.tex
\documentclass[paper]{subfiles}

Six-vertex models originate in statistical mechanics as a family of vertex models for crystal lattices with hydrogen bonds. Classically it is defined 
on a planar lattice region where each vertex of the lattice is connected by an edge to four \enquote{nearest neighbors}. A state of the model consists of an arrow on each edge such that the number of arrows pointing inwards at each vertex is exactly two. This 2-in-2-out law on the arrow configurations is called the \textit{ice rule}~\cite{doi:10.1063/1.1750821}. 
Thus there are six permitted types of local configurations around a vertex{\textemdash}hence the name \emph{six-vertex model} (see Figure~\ref{fig:orientations}). In 
graph theoretic terms, the states are Eulerian orientations of the underlying undirected graph.

\begin{figure}[h!]
\centering
\begin{subfigure}[b]{0.15\linewidth}
\centering\includegraphics[width=\linewidth]{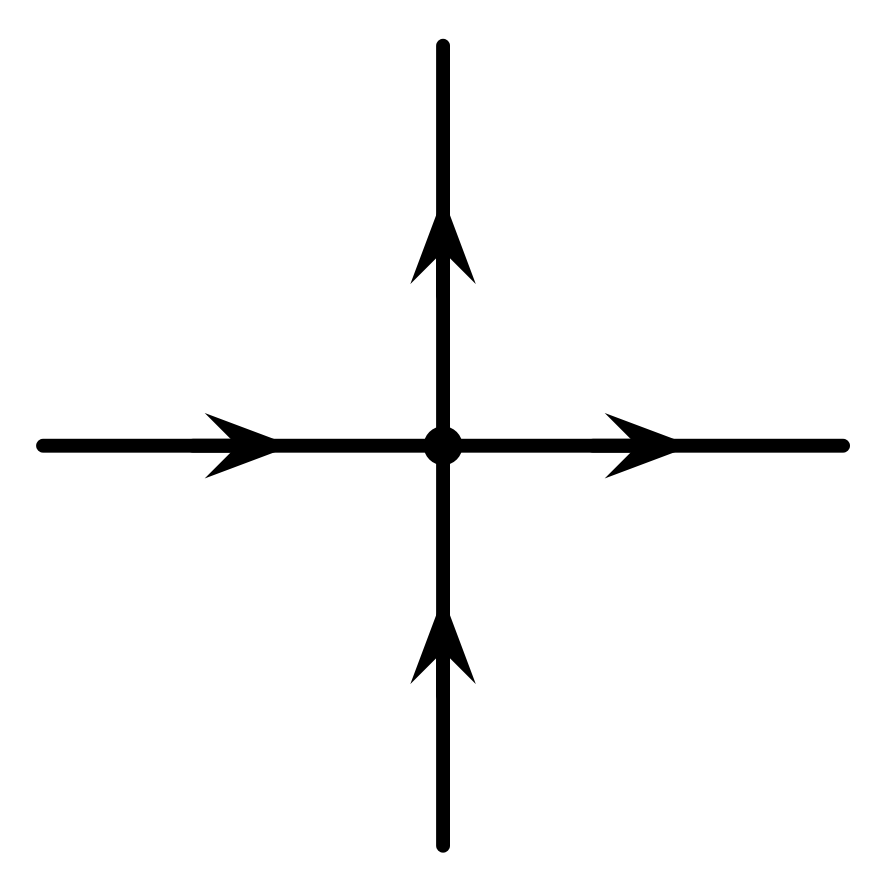}\caption{$1$}
\label{fig:orientations_1}
\end{subfigure}
\begin{subfigure}[b]{0.15\linewidth}
\centering\includegraphics[width=\linewidth]{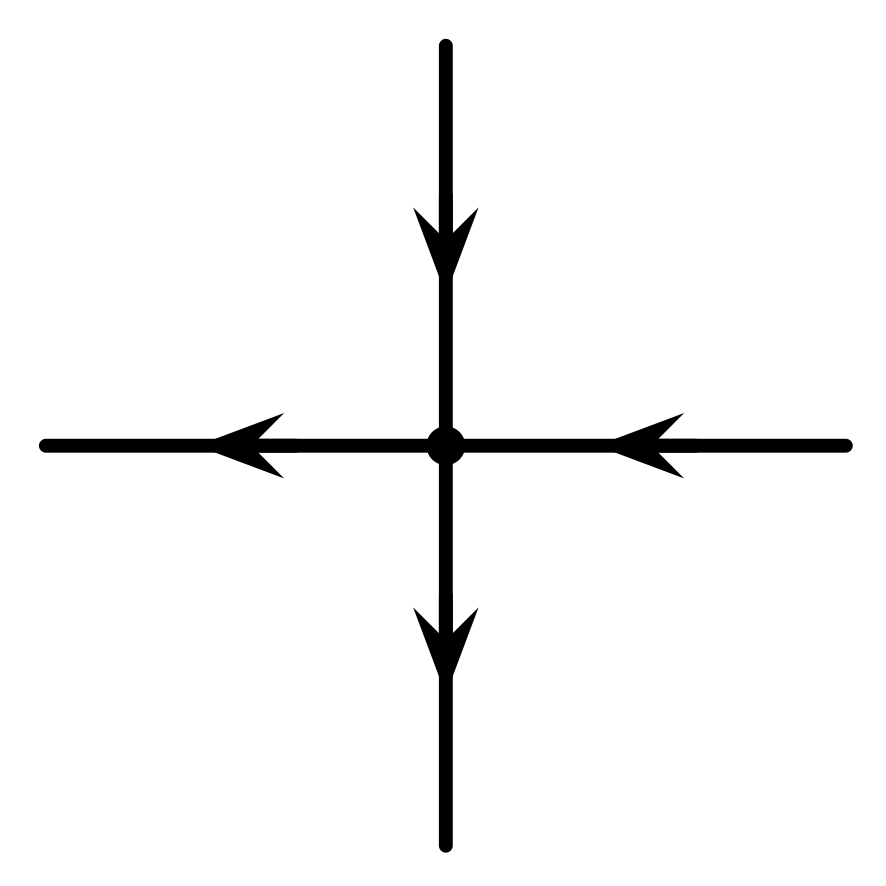}\caption{$2$}
\label{fig:orientations_2}
\end{subfigure}
\begin{subfigure}[b]{0.15\linewidth}
\centering\includegraphics[width=\linewidth]{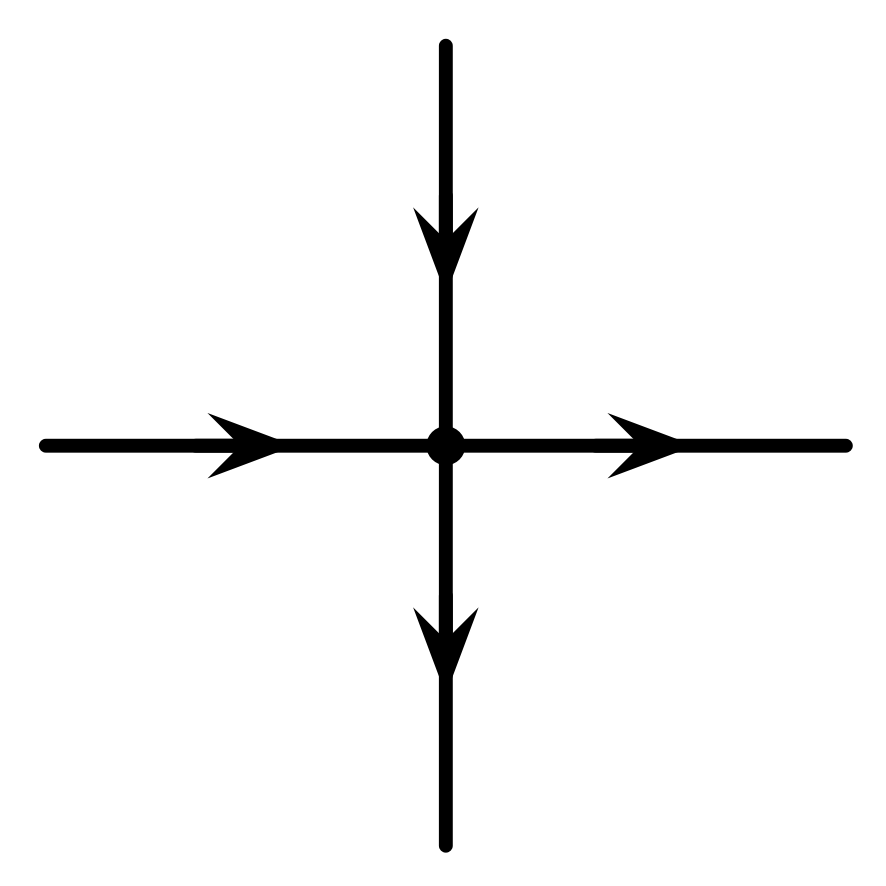}\caption{$3$}
\label{fig:orientations_3}
\end{subfigure}
\begin{subfigure}[b]{0.15\linewidth}
\centering\includegraphics[width=\linewidth]{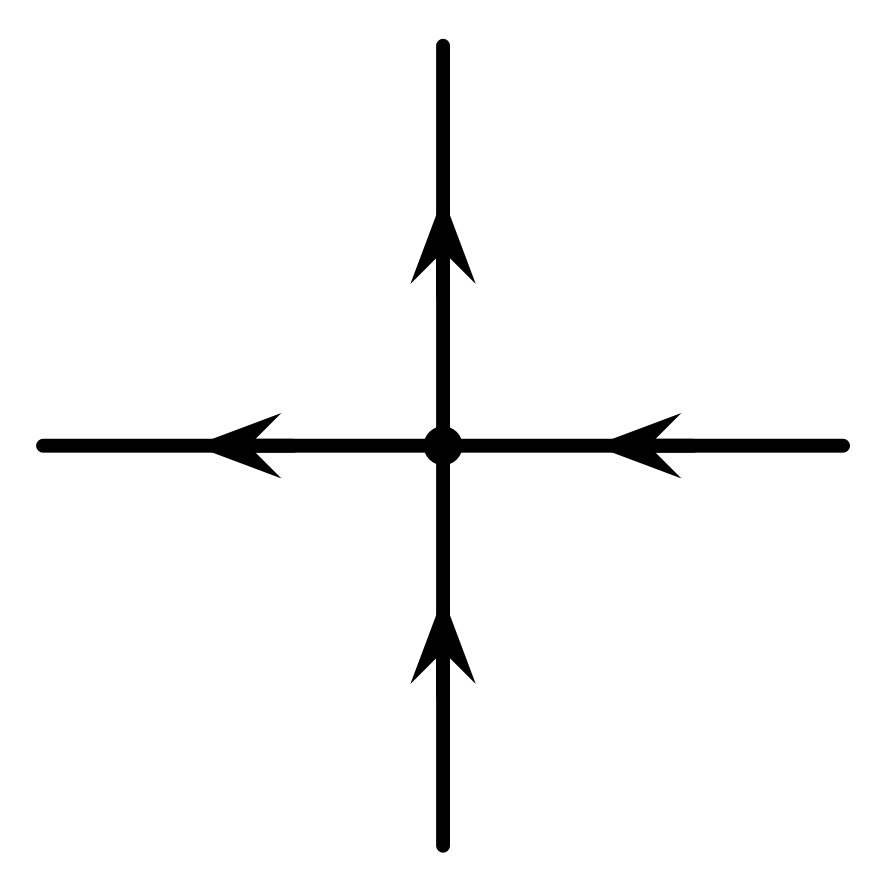}\caption{$4$}
\label{fig:orientations_4}
\end{subfigure}
\begin{subfigure}[b]{0.15\linewidth}
\centering\includegraphics[width=\linewidth]{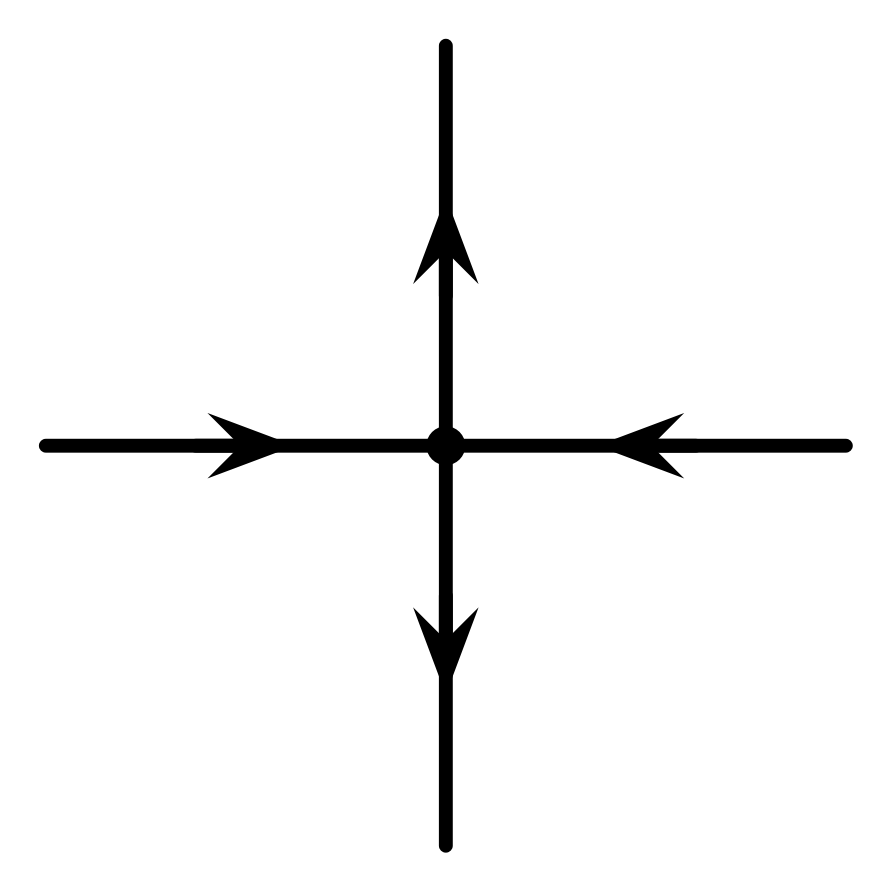}\caption{$5$}
\label{fig:orientations_5}
\end{subfigure}
\begin{subfigure}[b]{0.15\linewidth}
\centering\includegraphics[width=\linewidth]{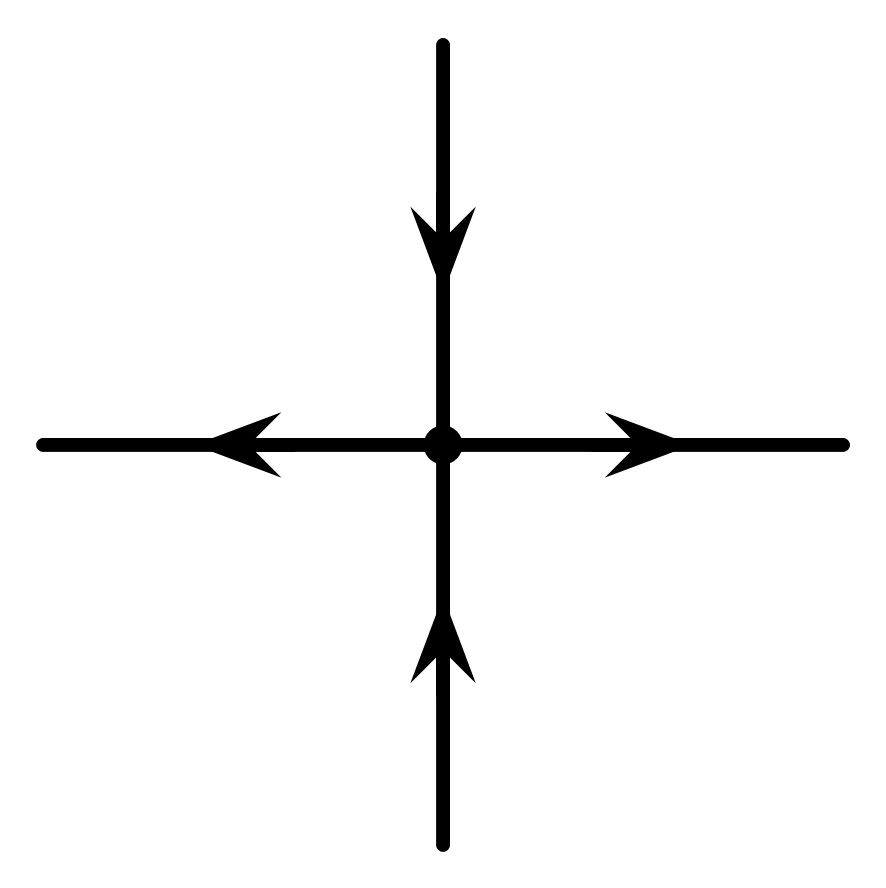}\caption{$6$}
\label{fig:orientations_6}
\end{subfigure}
\caption{Valid configurations of the six-vertex model.}\label{fig:orientations}
\end{figure}

In general, the six 
configurations 1 to 6 in Figure~\ref{fig:orientations}
are associated with six possible weights $w_1, \ldots, w_6$.
%
We will follow convention in physics and assume \textit{arrow reversal symmetry}\footnote{This is often assumed in physics. From Baxter's book \cite{Baxter:book}: \enquote{These ensure that on the square lattice the model is unchanged by reversing all arrows, which one would expect to be the situation for a model in zero external electric field. Thus this is a `zero-field' model which includes the ice, KDP and F models as special cases.}},
 i.e. $w_1 = w_2 = a, 
w_3 = w_4= b$ and $w_5 = w_6 = c$.
In this paper we assume $a, b, c \ge 0$, as is
assumed in classical physics.
 The partition function of the six-vertex model with parameters
$(a, b, c)$  on a 4-regular graph $G$, where incident edges of each vertex 
are labeled 1 to 4, is defined as
\[Z(G; a, b, c) = \sum_{\tau \in \mathcal{EO}(G)}a^{n_1 + n_2}b^{n_3 + n_4}c^{n_5 + n_6},\]
where $\mathcal{EO}(G)$ is the set of all Eulerian orientations of $G$,
 and $n_i$ is the number of vertices in type  $i$ ($1 \le i \le 6$) in the graph
under an Eulerian orientation $\tau \in \mathcal{EO}(G)$.

The first such models were introduced by Linus Pauling \cite{doi:10.1021/ja01315a102} in 1935 to describe the properties of ice. In 1967, Elliot Lieb~\cite{PhysRev.162.162, PhysRevLett.18.1046, PhysRevLett.19.108} 
 famously showed that,
for parameters $(a,b,c) = (1,1,1)$ on the square lattice graph,
as the side $N$ of the square approaches $\infty$,
the value of the \enquote{partition function per vertex}
$W = Z^{1/N^2}$ approaches $\left( \frac{4}{3} \right)^{3/2}
\approx 1.5396007\ldots$ (Lieb's square ice constant).
 This is called an exact solution of the model, and is considered a triumph.
 After that, exact solutions for other lattice type graphs (such as \cite{PhysRevLett.19.103, PhysRevB.2.723}) have been obtained in the limiting sense.

For half a century,
the six-vertex model has fascinated physicists, chemists,
mathematicians and others~\footnote{According to \textit{Google Scholar}, there are thousands of papers on the six-vertex model, comparable to that of ferromagnetic Ising and monomer-dimer models; these three
are the most studied models in statistical physics.}.
Beyond physics, connections of the six-vertex model to many other areas
 are discovered.
For example, 
Zeilberger~\cite{DBLP:journals/combinatorics/Zeilberger96}
 proved the famous alternating sign matrix (ASM)
conjecture in combinatorics, and Kuperberg \cite{doi:10.1155/S1073792896000128} gave a simplified proof making a connection to
 the six-vertex model.

The six-vertex model is also known to be
 related to the Tutte polynomial \cite{Ellis-Monaghan2011} 
in at least two points.
It is known~\cite{tutte54} that $T(G; 0, -2)$
is the number of Eulerian orientations,
i.e., $T(G; 0, -2) = Z(G; 1, 1, 1) = |\mathcal{EO}(G)|$,
 for every 4-regular graph $G$. 
Another link was proved by Las Vergnas \cite{VERGNAS1988367}
that $Z(H; 1, 1, 2) = 2T(G; 3,3)$ for any plane graph $G$ with medial graph $H$.

Recently, the exact computational complexity of six-vertex models
has been investigated.
This is studied in the context of a classification program for
the complexity of counting problems, where the six-vertex models
serve as important basic (asymmetric) cases
for Holant problems~\cite{Cai:journals/corr/CaiFS17}.
It is shown that there are some surprising P-time computable
settings, but for most parameters computing the partition
function $Z(G; a, b, c)$ exactly is \#P-hard.
 Under our parameterization of $a, b, c$ being nonnegative
(as is the case in the classical setting), the only 
P-time computable cases are: (1) two of $a, b, c$ are zero or (2) one of $a, b, c$ is zero and the other two are equal. Evaluation at any other point for a general graph
is \#P-hard. On planar graphs it is also P-time computable
for  parameter settings $(a,b,c)$ that
satisfy  $c^2 = a^2 + b^2$.
All other non-trivial  P-time computable cases require
cancellations (for real or complex parameters  $(a,b,c)$)
and do not apply for nonnegative $a, b, c$.
Mihail and Winkler first proved that
computing the number of unweighted Eulerian orientations
is \#P-complete over general graphs~\cite{Mihail1996}.
Huang and Lu proved that it remains  \#P-complete
for even degree regular (but
not necessarily planar) graphs~\cite{Huang:2016:DRW:2893602.2893654}.
Guo and Williams improved it to planar $4$-regular graphs~\cite{Guo2013}.
The latter is equivalent to computing the
 partition function of the six-vertex model on planar graphs
with the parameter setting $(1, 1, 1)$.


In terms of approximate complexity, results are limited.
To our best knowledge, there are only a very few papers that relate to
the approximate complexity of the six-vertex model, and they are all
on \emph{unweighted} Eulerian orientations.
Mihail and Winkler's pioneering work \cite{Mihail1996} gave the first \textit{fully polynomial randomized approximation scheme (FPRAS)} for the number of Eulerian orientations on a general graph. Luby, Randall, and Sinclair presented an elegant proof of the rapid mixing of a Markov chain that leads to a \textit{fully polynomial almost uniform sampler (FPAUS)} for Eulerian orientations on any region of the Cartesian lattice with fixed boundaries~\cite{doi:10.1137/S0097539799360355}. Randall and Tetali~\cite{doi:10.1063/1.533199} used a comparison technique to prove the single-site Glauber dynamics is rapidly mixing
on the same lattice graph,
by relating this Markov chain to the Luby-Randall-Sinclair chain.
Goldberg, Martin, and Paterson~\cite{RSA:RSA20002} further extended the 
technique by Randall and Tetali
to prove that the single-site Glauber dynamics is rapidly mixing 
for the free-boundary case on lattice graphs.

The known results on approximate complexity for the six-vertex model
are all for the unweighted case, which is the
point $(1, 1, 1)$ in the six-vertex model. 
In this paper we initiate a study toward a classification
of the approximate complexity of the six-vertex model
 in terms of the parameters. Our results conform
 to phase transitions in physics.

Here we briefly describe the phenomenon of phase transition of the 
zero-field six-vertex model (see Baxter's book~\cite{Baxter:book} for more details).
On square-lattice 
in the thermodynamic limit: (1) When  $a > b + c$ (FE: ferroelectric phase)
any finite region tends to be frozen into one of the two configurations 
where either all 
arrows point up or to the right~(Figure~\ref{fig:orientations}-1), or all point down or 
to the left~(Figure~\ref{fig:orientations}-2).
(2) Symmetrically when $b > a + c$ (also FE) all 
arrows point down or to the right~(Figure~\ref{fig:orientations}-3), or all point up or to the left~(Figure~\ref{fig:orientations}-4).
(3) When
$c > a + b$ (AFE: anti-ferroelectric phase)
configurations in Figure~\ref{fig:orientations}-5 and Figure~\ref{fig:orientations}-6 alternate.
(4) When $c < a + b$, $b < a + c$, and $a < b + c$, the system is disordered 
(DO: disordered phase)
 in the sense that all correlations decay to zero
with increasing distance; in particular
on the dashed curve $c^2 = a^2 + b^2$ the model can be solved by Pfaffians
exactly~\cite{PhysRevB.2.723}, and the correlations decay 
inverse polynomially, rather than exponentially, in distance.
 See Figure~\ref{fig:transition_phase}.

\captionsetup[subfigure]{labelformat=parens}
\begin{figure}[h!]
\centering
\begin{subfigure}[t]{0.48\linewidth}
\centering\includegraphics[width=\linewidth]{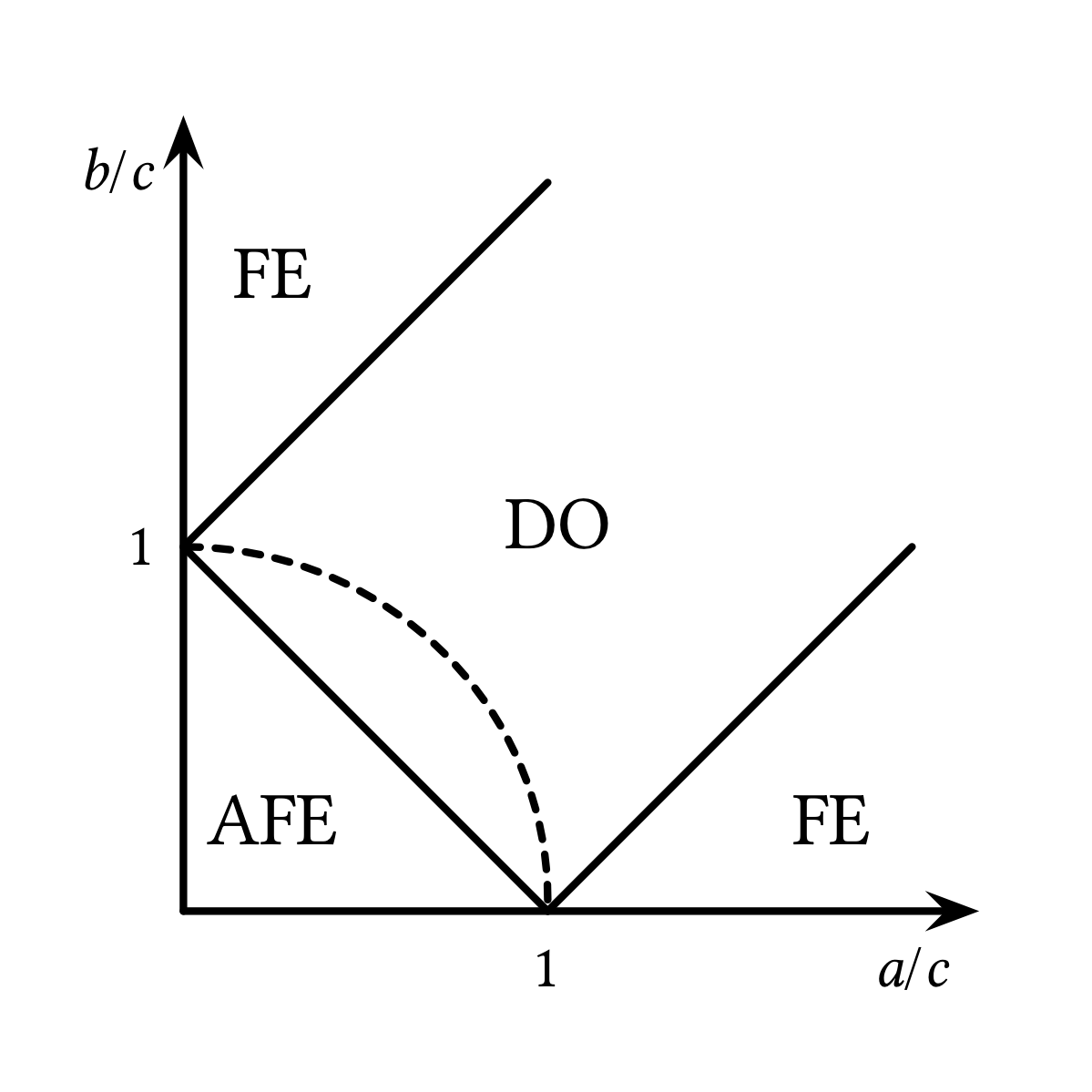}
\caption{Phase diagram of the six-vertex model.}
\label{fig:transition_phase}
\end{subfigure}
\hfill
\begin{subfigure}[t]{0.48\linewidth}
\centering\includegraphics[width=\linewidth]{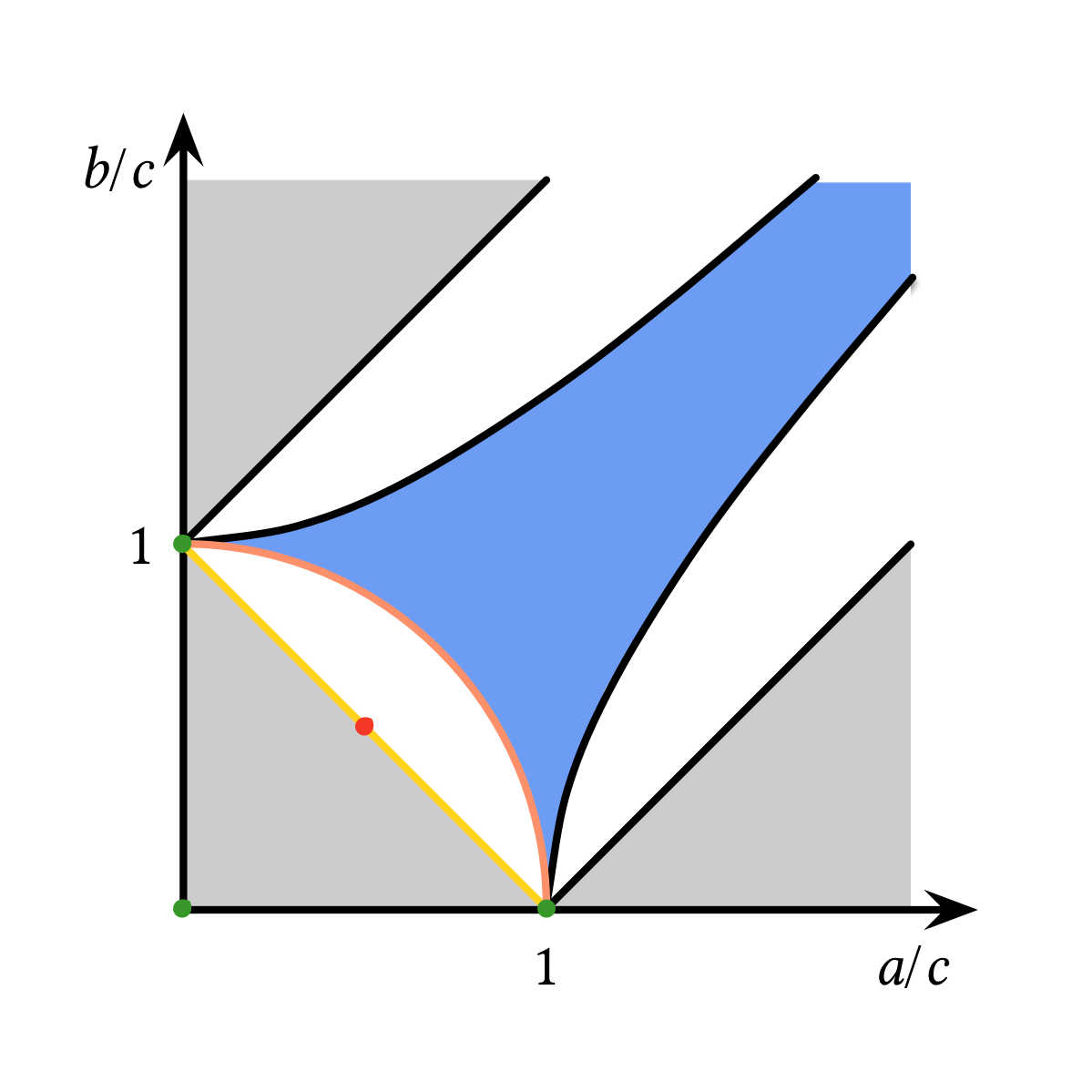}
\caption{Complexity diagram of the six-vertex model.}
\label{fig:transition_complexity}
\end{subfigure}
\caption{}\label{fig:transitions}
\end{figure}

In Figure~\ref{fig:transition_complexity} we have a corresponding
complexity landscape.
\begin{thm}\label{results}
There is an FPRAS for $Z(G; a,b,c)$
if  $a^2 \le b^2 + c^2$, $b^2 \le a^2 + c^2$, and $c^2 \le a^2 + b^2$
(the blue region).
There is no FPRAS for  $Z(G; a,b,c)$
if 
$a > b + c$ or $b > a + c$ or $c > a + b$
(the grey region), unless RP $=$ NP.
\end{thm}

Our FPRAS result is actually stronger in that
the FPRAS works even if different signatures from the blue region are assigned
at different vertices.
The blue region is a proper subset of the disordered phase.
The point $(1,1,1)$ is contained in this region,
which is the only previously known approximable case.
The hardness part (the grey region) coincides with the 
FE/AFE phases.
The three green points together with a point at infinity 
($(a,b,c) = (1,1,0)$)
are exactly P-time computable.
 All parameters belonging to the orange curve $c^2 = a^2 + b^2$ are
 exactly P-time computable on planar graphs.
Computing for the six-vertex model at $(1/2, 1/2, 1)$
(the red point) is equivalent to evaluating
the Tutte polynomial $T(G; 3,3)$ on planar graphs.
Note that any 4-regular plane graph $H$ is the medial graph of some plane graph $G$.
The approximation complexity for the
white region
is unknown.

Furthermore, we show that there is a fundamental
structural difference in the behavior on the  two sides separated by the phase transition threshold, in terms of closure properties.  \textit{Gadget construction} is a common technique used in  approximation-preserving reductions~\cite{Dyer2004}. If a constraint function $g$ can be expressed by a polynomial-size 
gadget using a constraint function $f$,
then the approximation complexity of $g$ is no harder than that of $f$.
In  \thmref{conf_1} of \secref{sec:theorems}, we prove that
the set of 4-ary functions lying in the combined region of blue and
white (this is the same as the DO region in~Figure~\ref{fig:transition_phase})
is closed under gadget construction.
In \thmref{conf_2}
we prove that the set of 4-ary functions lying on the yellow line (phase transition threshold for AFE and DO) is closed under \textit{planar} gadget construction.
\thmref{conf_1} is also used in proving a Markov chain is rapidly mixing in \secref{sec:fpras}.



Our FPRAS also has implications for counting weighted sum of
directed Eulerian partitions (partition
 of edges of $G$ into directed edge-disjoint circuits).
A special case is an FPRAS for this weighted sum 
when the weight of \sub{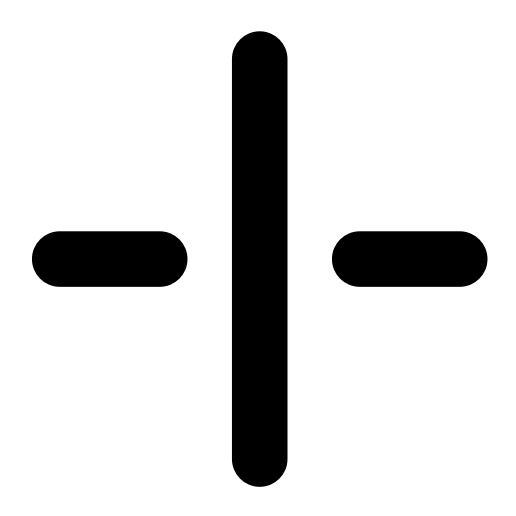} is at least $\sqrt{2} -1$ (more on the
connection between directed Eulerian partitions and the 
three types of pairings \sub{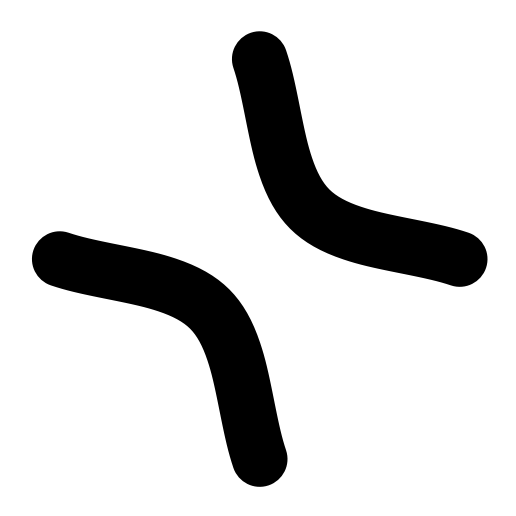}, \sub{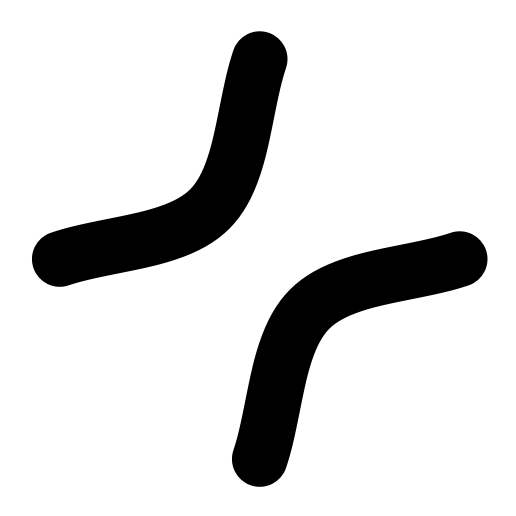}, and \sub{3}
can be found in \secref{sec:theorems}).

Our proof uses the Holant framework.
In \secref{sec:prelim} we express the six-vertex model as
a Holant problem. This allows us to use techniques developed in
the study of Holant problems to make progress in  both fronts:
We design a rapidly mixing Markov chain to derive a  
FPRAS in the blue region (within the disordered
phase). This result can also be obtained by
 using a technique called \emph{windable} by McQuillan~\cite{DBLP:journals/corr/abs-1301-2880}, specifically developed for the Holant framework.
 We also use techniques developed in the
Holant framework  to prove NP-hardness of approximation for
the six-vertex model in the grey
region (coincide with the ferroelectric/anti-ferroelectric phases).
These are the first inapproximability results
for the six-vertex model.


%% file: prelim.tex
\documentclass[paper]{subfiles}

\subsection{Six-Vertex Model as a Holant Problem}
The six-vertex model is naturally expressed as a Holant problem, which
we define as follows.
 A function $f: \left\{0,1\right\}^k \rightarrow \mathbb{C}$ is
called a constraint function, or a \textit{signature}, of arity $k$.
In this paper we restrict $f$ to take nonnegative values in  $\mathbb{R}^+$.
 Fix a set $\mathcal{F}$ of constraint functions. 
A signature grid $\Gamma = (G, \xi)$ is a tuple, where $G = (V, E)$ is a graph, $\xi$ labels each $v \in V$ with a function $f_v \in \mathcal{F}$ of arity $\operatorname{deg}(v)$, and the incident edges $E(v)$ at $v$ are
identified as input variables to $f_v$, also labeled by $\xi$.
Every assignment $\sigma: E \rightarrow \left\{0, 1\right\}$ gives
an evaluation  $\prod_{v \in V}f_v\left(\sigma |_{E(v)}\right)$, 
where $\sigma |_{E(v)}$ denotes the restriction of $\sigma$ to $E(v)$. 
The problem  $\Holant\left(\mathcal{F}\right)$ on an instance $\Gamma$ is to compute
$\Holant\left(\Gamma; \mathcal{F}\right) = \sum_{\sigma:E\rightarrow\left\{0,1\right\}}\prod_{v\in V}f_v\left(\sigma |_{E(v)}\right)$.
When $\mathcal{F} = \left\{f\right\}$ is a singleton set, we write $\Holant\left(f\right)$ for simplicity. We use $\Holant\left(\mathcal{F} | \mathcal{G}\right)$ for Holant problems over signature grids with a bipartite graph $(U, V, E)$ where each vertex in $U$ (or $V$) is assigned a signature in $\mathcal{F}$ (or $\mathcal{G}$, respectively).

To write the six-vertex model on a 4-regular graph $G = (V, E)$ as a Holant problem,
consider the edge-vertex incidence graph $G' = (U_E, U_V, E')$ of $G$.
We model the orientation of an edge in $G$ 
by putting the $\textsc{Disequality}$ signature $(\neq_2)$ (which outputs 1 on inputs 01, 10 and outputs 0 on 00, 11)
on  $U_E$ in $G'$.
 We say an orientation on edge $e = \left\{w, v\right\} \in E$ is going out $w$ and into 
 $v$ in $G$ if the edge $(u_e, u_w) \in E'$ in $G'$
 takes value $1$ (and  $(u_e, u_v) \in E'$ takes value $0$).
An arity-4 signature $f$ on input $x_1, x_2, x_3, x_4$ has the \textit{signature matrix} $M(f) = M_{x_1x_2,x_4x_3}(f) = \left[\begin{smallmatrix} f_{0000} & f_{0010} & f_{0001} & f_{0011} \\ f_{0100} & f_{0110} & f_{0101} & f_{0111} \\ f_{1000} & f_{1010} & f_{1001} & f_{1011} \\ f_{1100} & f_{1110} & f_{1101} & f_{1111}\end{smallmatrix}\right]$, where the order reversal $x_4x_3$ is for the 
convenience of composing these signatures in a planar fashion,
by matrix product. At a vertex  $v$ in $G$, if we locally index the left, down, right, and up edges incident to $v$ by 1, 2, 3, and 4, respectively, then the 
constraint by the six-vertex model as specified in \figref{fig:orientations} can be expressed perfectly as a signature $f$ with $M(f) = \left[\begin{smallmatrix} 0 & 0 & 0 & a \\ 0 & b & c & 0 \\ 0 & c & b & 0 \\ a & 0 & 0 & 0\end{smallmatrix}\right]$.
Thus computing the partition function $Z(G; a, b, c)$ is equivalent to evaluating $\Holant\left(G'; \neq_2 | f\right)$ for this $f$.

For convenience in presenting our theorems and proofs, we adopt the following notations assuming $a, b, c \in \mathbb{R}^+$.
We assume $f$ has signature matrix  $M(f) = \left[\begin{smallmatrix} & & & a \\ & b & c & \\ & c & b & \\ a & & &\end{smallmatrix}\right]$.
\setlist[itemize]{itemsep=-1mm}
\begin{itemize}
\item
$\mathcal{F}_{\le^2} := \{f \; | \; 
a^2 \le b^2 + c^2,~~ b^2 \le a^2 + c^2,~~ c^2 \le a^2 + b^2\}$;
\item
$\mathcal{F}_\le ~:= \{f \; | \; 
a \le b + c,~~ b \le a + c,~~ c \le a + b\}$;
\item
$\mathcal{F}_= ~:= \{f \; | \; 
c = a + b\}$;
\item
$\mathcal{F}_> ~:= \{f \; | \; 
a > b + c ~\text{ or }~ b > a + c ~\text{ or }~ c > a + b \text{ where } a, b, c > 0\}$.
\end{itemize}
\begin{rem}
$\mathcal{F}_{\le^2} \subset \mathcal{F}_\le$.
\end{rem}

\subsection{Approximation Algorithms}
If a counting problem is \#P-hard, 
we may still hope that the problem can be approximated. Suppose $f: \Sigma^* \rightarrow \mathbb{R}$ is a function mapping problem instances to real numbers. A \textit{fully polynomial randomized approximation scheme (FPRAS)} \cite{Karp:1983:MAE:1382437.1382804} for a problem is a randomized algorithm that takes as input an instance $x$ and $\varepsilon > 0$, running in time polynomial in $n$ (the input length) and $\varepsilon^{-1}$, and outputs a number $Y$ (a random variable) such that
\[\operatorname{Pr}\left[(1 - \varepsilon)f(x) \le Y \le (1 + \varepsilon)f(x)\right] \ge \frac{3}{4}.\]
%

%% file: theorems.tex
\documentclass[paper]{subfiles}
\begin{thm}\label{conf_1}
If $f$ is the signature of a $4$-ary gadget on the right hand side of $\Holant\left(\neq_2 | \mathcal{F}_\le\right)$, then $f \in \mathcal{F}_\le$.
\end{thm}
\begin{thm}\label{conf_2}
If $f$ is the signature of a $4$-ary plane gadget on the right hand side of $\Holant\left(\neq_2 | \mathcal{F}_=\right)$, then $f \in \mathcal{F}_=$.
\end{thm}

Before proving \thmref{conf_1} and \thmref{conf_2}, we introduce another view of the six-vertex model. A valid configuration in the six-vertex model, i.e. a weighted Eulerian orientation, can also be viewed as a combination of weighted directed Eulerian partitions.
An \textit{Eulerian partition} of a graph $G$ is a partition of the edges of $G$ into edge-disjoint \textit{circuits} (in which vertices may repeat whereas edges cannot). A \textit{directed Eulerian partition} is an Eulerian partition where every edge-disjoint circuit takes one of the two cyclic orientations.
Let $G = (V, E)$ be a 4-regular graph and $v$ be a vertex of $G$. Let $e_1, e_2, e_3, e_4$ be the four edges incident to $v$. A \textit{pairing} $\varrho$ at $v$ is a partition of $\{e_1, e_2, e_3, e_4\}$ into pairs. There are exactly three distinct pairings at $v$~(\figref{fig:pairings}) which we denote by three special symbols: $\sub{1}, \sub{2}, \sub{3}$, respectively.
An Eulerian partition of $G$ can be uniquely determined by a family of pairings $\varphi = \{\varrho_v\}_{v \in V}$, where $\varrho_v \in \{\sub{1}, \sub{2}, \sub{3}\}$ is a pairing at $v${\textemdash}once the pairing at each vertex is fixed, then the two edges paired together at each vertex is also adjacent in the same circuit.

\begin{figure}[h!]
\centering
\begin{subfigure}[b]{0.3\linewidth}
\centering\includegraphics[width=0.5\linewidth]{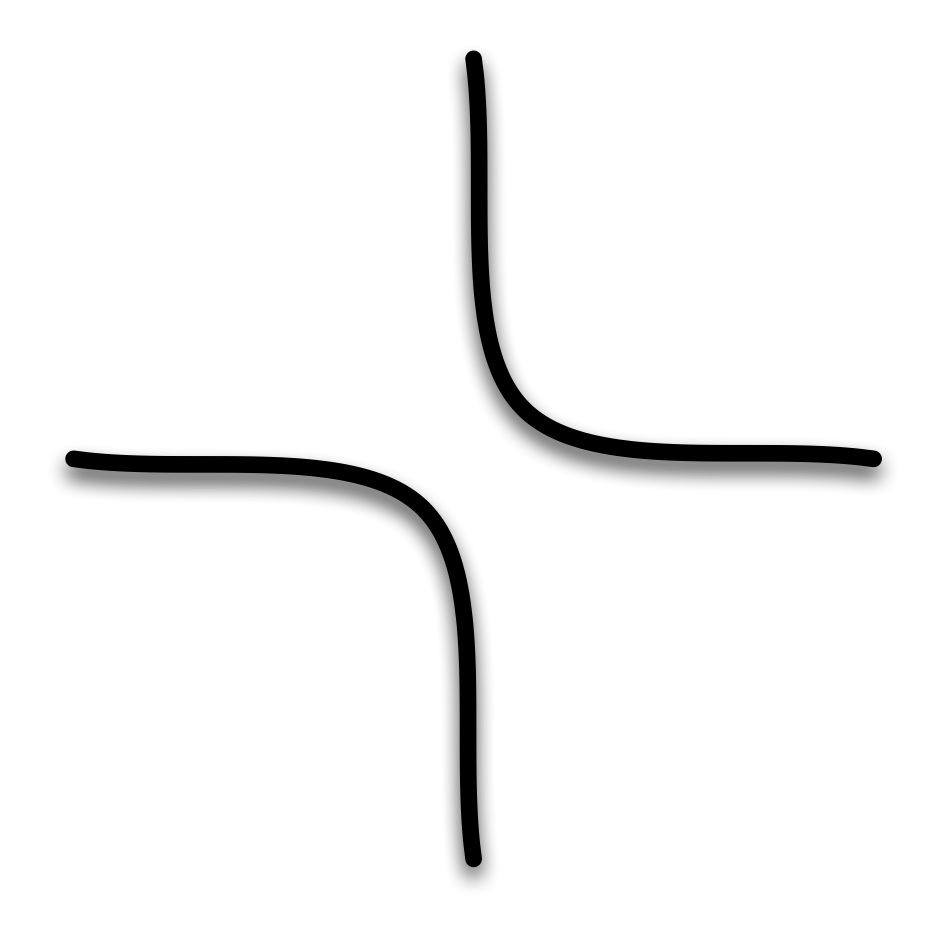}\caption{$\sub{1}$}
\end{subfigure}
\begin{subfigure}[b]{0.3\linewidth}
\centering\includegraphics[width=0.5\linewidth]{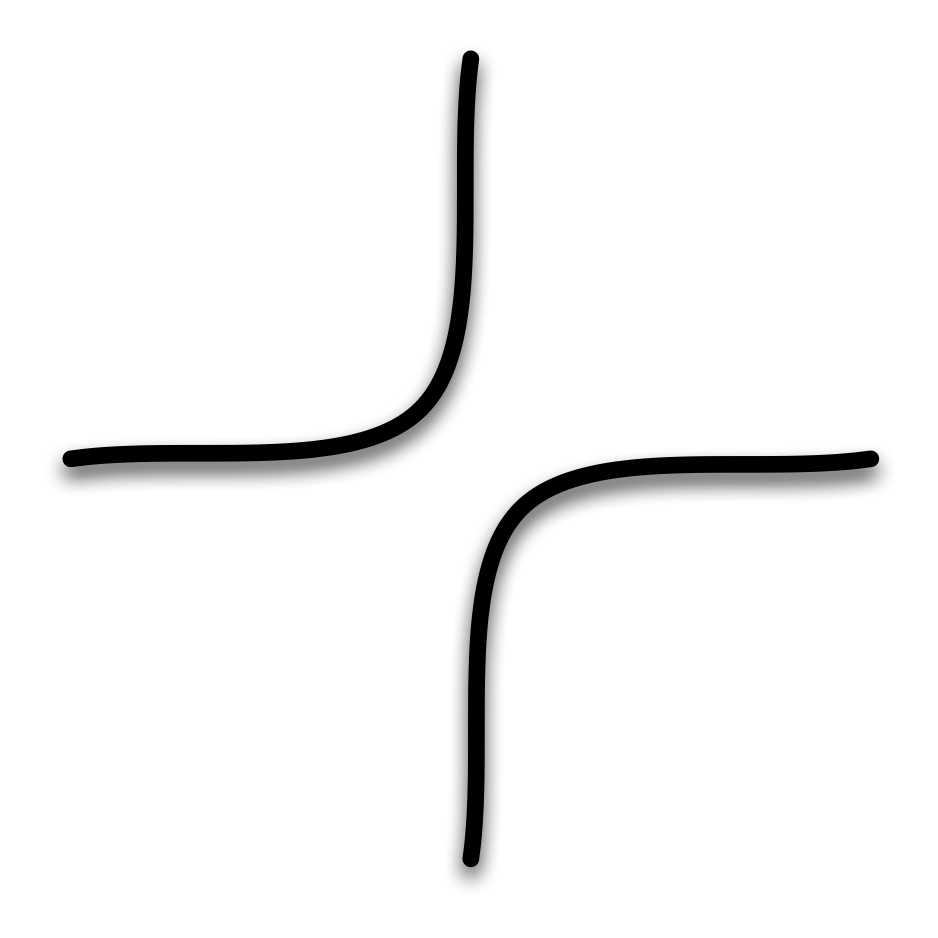}\caption{$\sub{2}$}
\end{subfigure}
\begin{subfigure}[b]{0.3\linewidth}
\centering\includegraphics[width=0.5\linewidth]{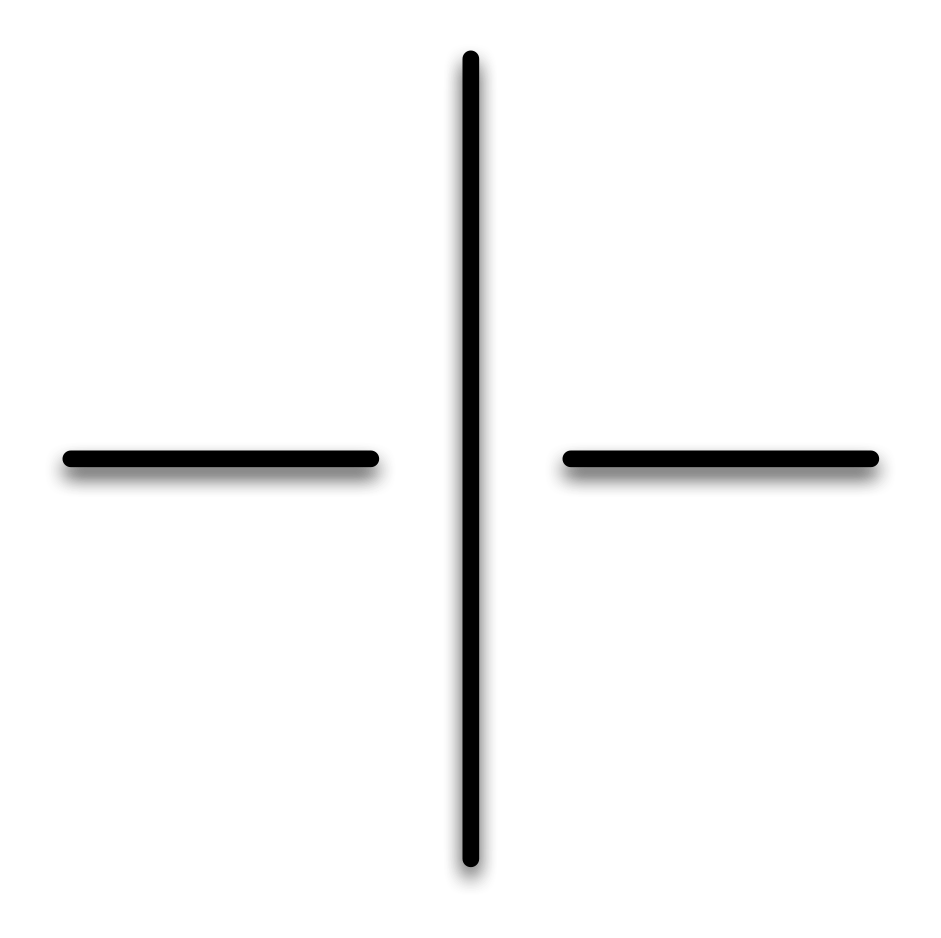}\caption{$\sub{3}$}
\end{subfigure}  
\caption{Pairings at a degree 4 vertex.}\label{fig:pairings}
\end{figure}

For any vertex $v$ in a valid configuration $\tau$ of the six-vertex model (where ice rule is satisfied), incoming edges can be paired with outgoing edges in exactly two ways, corresponding to two of the three pairings at $v$. For example, the configuration in \figref{fig:orientations}-1 of the six-vertex model has two underlying pairings, $\sub{2}$ and $\sub{3}$. Therefore, $\tau$ can be decomposed into $2^{|V|}$ distinct directed Eulerian partitions denoted by $\Phi(\tau)$.
Since no two Eulerian orientations share one directed Eulerian partition and every directed Eulerian partition corresponds to a particular Eulerian orientation, the map from six-vertex configurations to directed Eulerian partitions is $1$-to-$2^{|V|}$, non-overlapping, and surjective.
Define $w$ to be a function assigning a weight to every pairing at every vertex and let the weight $\tilde{w}(\varphi)$ of an Eulerian partition $\varphi$, undirected or directed, be the product of weights at each vertex.
In particular, when $w$ is defined such that
$\left\{\begin{smallmatrix}
w(\subinmatrix{1}) = \frac{-a + b + c}{2} \\
w(\subinmatrix{2}) = \frac{a - b + c}{2} \\
w(\subinmatrix{3}) = \frac{a + b - c}{2}
\end{smallmatrix}\right.$,
or equivalently
$\left\{\begin{smallmatrix}
a = w(\subinmatrix{2}) + w(\subinmatrix{3}) \\
b = w(\subinmatrix{1}) + w(\subinmatrix{3}) \\
c = w(\subinmatrix{1}) + w(\subinmatrix{2})
\end{smallmatrix}\right.$,
for every vertex with signature matrix $\left[\begin{smallmatrix} & & & a \\ & b & c & \\ & c & b & \\ a & & &\end{smallmatrix}\right]$,
then the weight of a six-vertex model configuration $\tau$ is equal to $\sum_{\varphi \in \Phi(\tau)}\tilde{w}(\varphi)$, by expressing a product of sums as a sum of products.

The connection between Eulerian orientations and Eulerian partitions on 4-regular graphs has been explored. Las Vergnas \cite{VERGNAS1988367} demonstrated a special case for plane graphs: the number of directed \textit{non-intersecting} Eulerian partitions is equal to the number of Eulerian orientations with weight 2 on every \textit{saddle} configuration (\figref{fig:orientations}-5~\ref{fig:orientations}-6), which is the six-vertex model at (1, 1, 2) .
Jaeger \cite{Jaeger1990} proposed a graph polynomial called \textit{transition polynomial} as a generalization of weighted Eulerian partitions, and related it with weighted Eulerian orientations.
The idea of unweighted directed Eulerian partitions was implicitly used in Mihail and Winkler's paper \cite{Mihail1996} to approximate the number of unweighted Eulerian orientations, where they also adopted the notion of \textit{pairings}.

\begin{proof}[Proof of \thmref{conf_1}]
For the signature $f$ of a 4-ary gadget on the right hand side of $\Holant\left(\neq_2 | \mathcal{F}_\le\right)$ (\figref{fig:gadget_plain}), we first show that its signature matrix must be of the form $\left[\begin{smallmatrix} & & & a' \\ & b' & c' & \\ & c' & b' & \\ a' & & &\end{smallmatrix}\right]$.
First, $f$ still obeys the ice rule, i.e. it cannot take nonzero values on inputs with Hamming weight not 2.
Including the dangling edges, 
every vertex has exactly two incoming
edges and two outgoing edges. 
Thus if we sum the in-degrees over all vertices,
it must equal to the
sum of out-degrees over all vertices,
i.e., $\sum_v {\rm in\text{-}deg}(v) 
 = \sum_v {\rm out\text{-}deg}(v)$.
Every internal edge contributes
exactly 1 to each sum. Thus the number of
incoming dangling edges is equal to
the number of
outgoing dangling edges, which must be 2
each since they sum to 4.
Second, $f$ still satisfies arrow reversal symmetry. For any valid orientation of edges in the gadget contributing a nonnegative factor to $f(x)$,
reversing the orientations on all edges will contribute the same factor to $f(\overline{x})$, as is true for every signature on a single vertex of degree 4.

\captionsetup[subfigure]{labelformat=parens}
\begin{figure}[h!]
\centering
\begin{subfigure}[b]{0.32\linewidth}
\centering\includegraphics[width=0.8\linewidth]{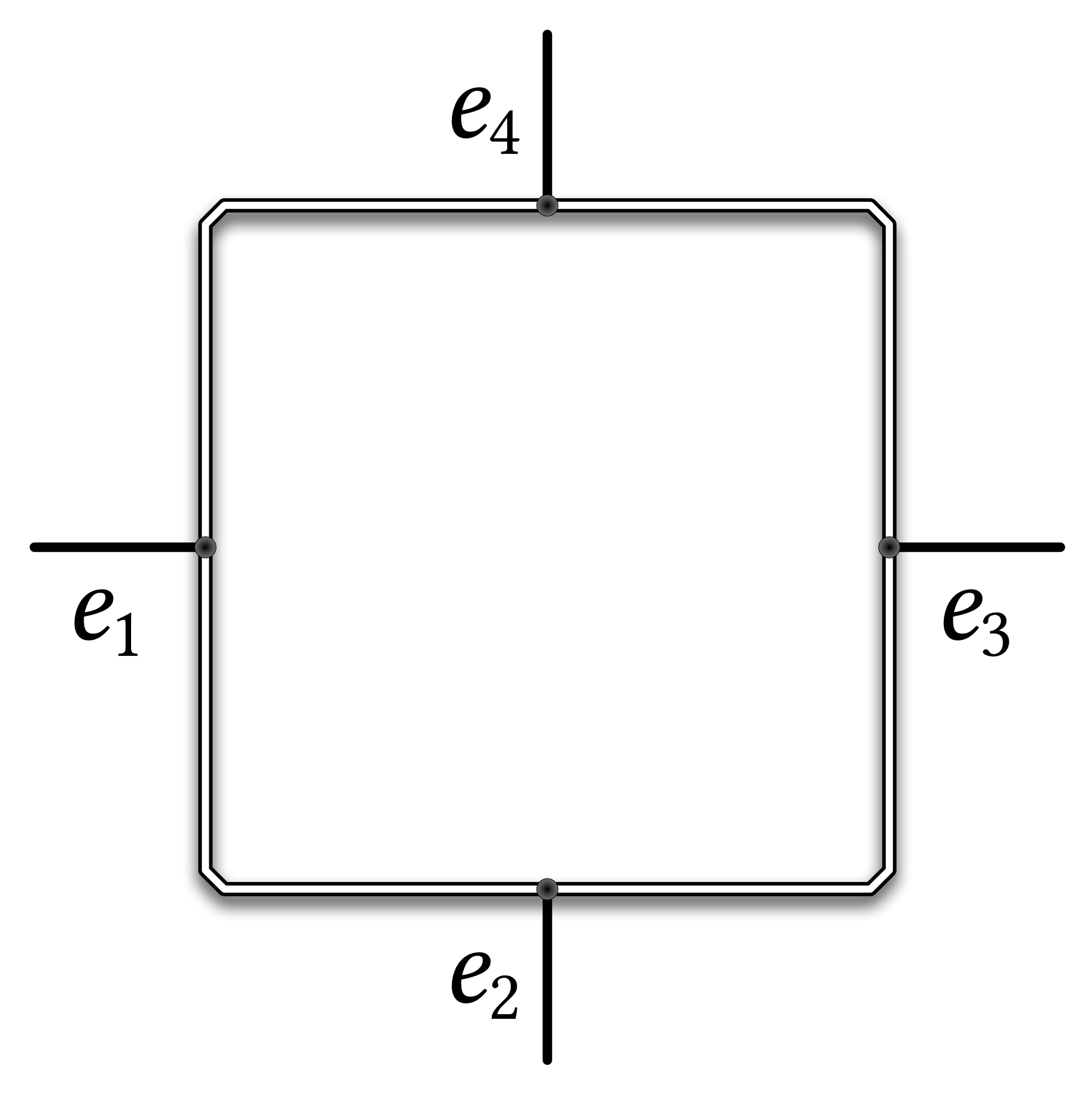}\caption{}\label{fig:gadget_plain}
\end{subfigure}
\begin{subfigure}[b]{0.32\linewidth}
\centering\includegraphics[width=0.8\linewidth]{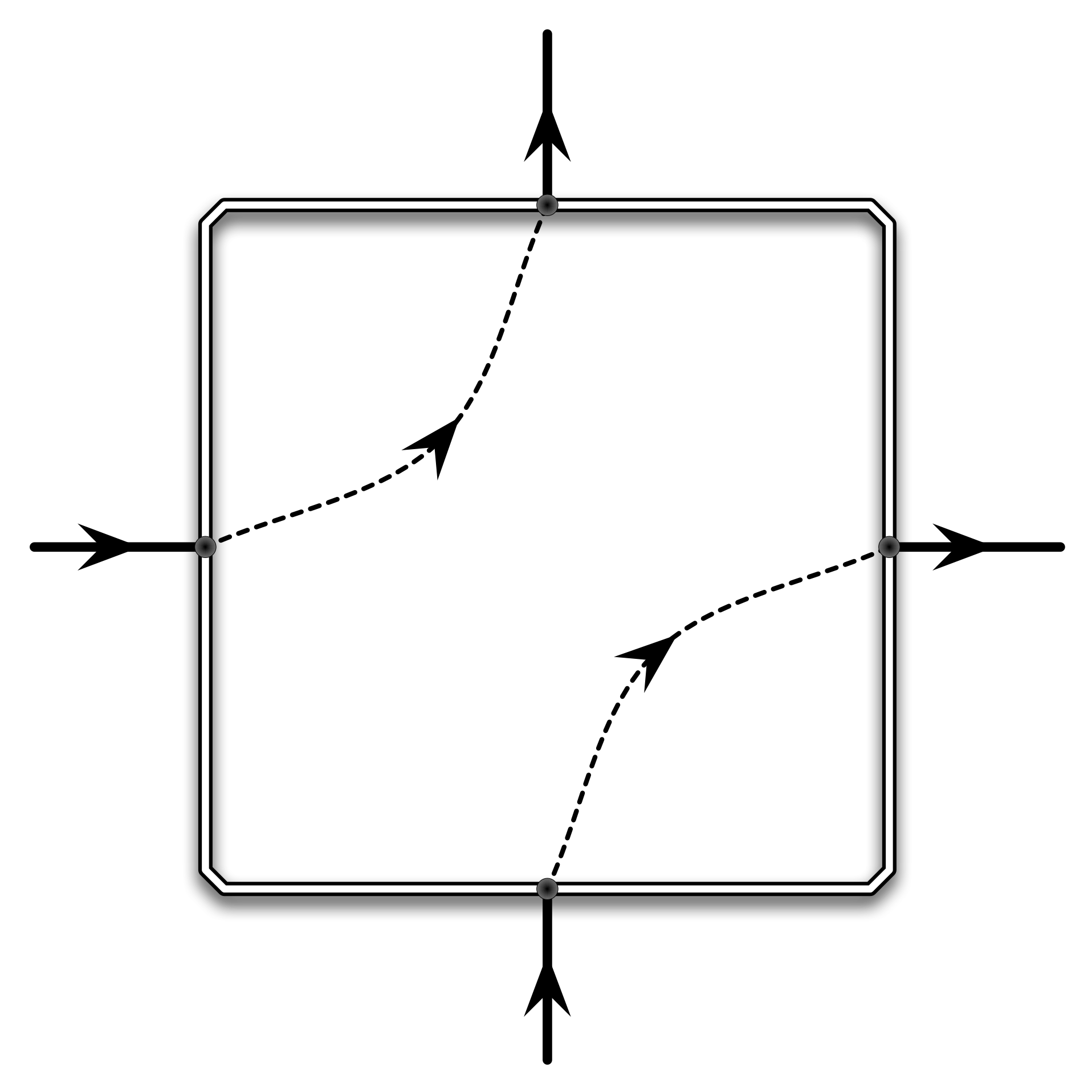}\caption{}\label{fig:gadget_a}
\end{subfigure}
\begin{subfigure}[b]{0.32\linewidth}
\centering\includegraphics[width=0.8\linewidth]{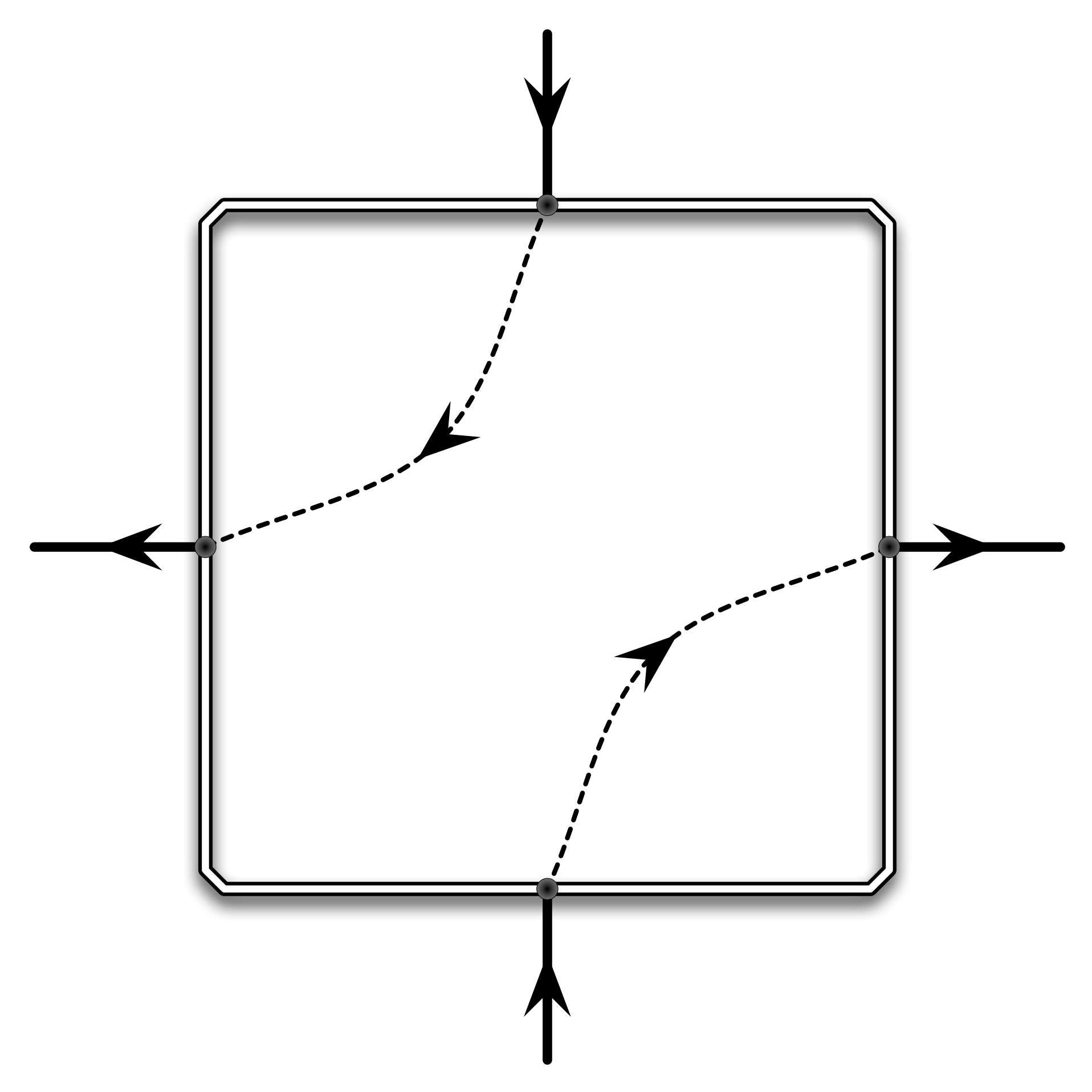}\caption{}\label{fig:gadget_c}
\end{subfigure}  
\caption{A gadget constructed on the right hand side of $\Holant\left(\neq_2 | \mathcal{F}_\le\right)$.}\label{fig:gadget}
\end{figure}

The notion of Eulerian partitions previously used for graphs can also be defined for gadgets. An \textit{Eulerian partition for a gadget} $g$ with four dangling edges is a partition of the edges in $g$ into edge-disjoint circuits and exactly two \textit{walks} (in which vertices may repeat whereas edges cannot) whose ends are exactly the four dangling edges. The weight $\tilde{w}$ of such an Eulerian partition $\varphi$ can be similarly defined. Set $w$ such that
$\left\{\begin{smallmatrix}
w(\subinmatrix{1}) = \frac{-a + b + c}{2} \\
w(\subinmatrix{2}) = \frac{a - b + c}{2} \\
w(\subinmatrix{3}) = \frac{a + b - c}{2}
\end{smallmatrix}\right.$,
or equivalently
$\left\{\begin{smallmatrix}
a = w(\subinmatrix{2}) + w(\subinmatrix{3}) \\
b = w(\subinmatrix{1}) + w(\subinmatrix{3}) \\
c = w(\subinmatrix{1}) + w(\subinmatrix{2})
\end{smallmatrix}\right.$.
Observe that if a vertex has a signature $f \in \mathcal{F}_\le$, then the weight of every pairing is \textit{nonnegative}, and the weight of any directed Eulerian partition of a graph/gadget comprised of such vertices is also \textit{nonnegative}.

Under the six-vertex model, for any specific configuration $\tau$ of the gadget with signature $f$ that contributes a nonzero factor to $f(0011)$ when $e_1, e_2$ go in and $e_3, e_4$ go out, it can be viewed as a weighted sum of directed Eulerian partitions $\Phi(\tau)$. For every Eulerian partition $\varphi \in \Phi(\tau)$, the two directed walks are either $\{e_1 \leadsto e_4, e_2 \leadsto e_3\}$ (\figref{fig:gadget_a}) or $\{e_1 \leadsto e_3, e_2 \leadsto e_4\}$.
Denote by $\Phi_{0011,\sub{2}}$ the set of directed Eulerian partitions (distributed in potentially many different six-vertex configurations), each of which has directed walks $\{e_1 \leadsto e_4, e_2 \leadsto e_3\}$; denote by $\Phi_{0011,\sub{3}}$ the set of directed Eulerian partitions, each of which has directed walks $\{e_1 \leadsto e_3, e_2 \leadsto e_4\}$.
In terms of directed Eulerian partitions of the gadget, $f(0011)$ can be seen as the weighted sum of elements from two disjoint sets $\Phi_{0011,\sub{2}}$ and $\Phi_{0011,\sub{3}}$.
Defining the weight of a set $\Phi$ of directed Eulerian partitions by $W(\Phi) = \sum_{\varphi \in \Phi} \tilde{w}(\varphi)$ yields $f(0011) = W(\Phi_{0011,\sub{2}}) + W(\Phi_{0011,\sub{3}})$, and similarly $f(1100) = W(\Phi_{1100,\sub{2}}) + W(\Phi_{1100,\sub{3}})$.
Note that there is a bijective weight-preserving map between $\Phi_{0011,\sub{2}}$ and $\Phi_{1100,\sub{2}}$ by reversing the direction of every circuit and walk of an Eulerian partition. That is to say, $W(\Phi_{0011,\sub{2}}) = W(\Phi_{1100,\sub{2}})$ and similarly $W(\Phi_{0011,\sub{3}}) = W(\Phi_{1100,\sub{3}})$.
This proves that $f(0011) = f(1100)$, as we noted earlier. Similar conclusions can be made for the other two pairs of values $\{f(0110), f(1001)\}$ and $\{f(0101), f(1010)\}$.

An important observation is that for each Eulerian partition in $\Phi_{0011,\sub{2}}$, if we only reverse the walk from $e_1 \leadsto e_4$ to $e_4 \leadsto e_1$ and keep the directions on all circuits and the other walk unchanged, this Eulerian partition has the same weight but now lies in $\Phi_{1010,\sub{2}}$ (\figref{fig:gadget_c}).
This is because at every vertex $v$, reversing any orientation of a branch of the given 
pairing $\varrho_v \in \{\sub{1}, \sub{2}, \sub{3}\}$ does not change the value $w(\varrho_v)$.
In this way, we set up a one-to-one weight-preserving map between $\Phi_{0011,\sub{2}}$ and $\Phi_{1010,\sub{2}}$, i.e. $W(\Phi_{0011,\sub{2}}) = W(\Phi_{1010,\sub{2}})$. Combining the result in the last paragraph, we can write
\setlist[itemize]{itemsep=-1mm}
\begin{itemize}
\item
$W(\sub{2}) = W(\Phi_{0011,\sub{2}}) = W(\Phi_{1100,\sub{2}}) = W(\Phi_{0101,\sub{2}}) = W(\Phi_{1010,\sub{2}})$;
\item
$W(\sub{1}) = W(\Phi_{0110,\sub{1}}) = W(\Phi_{1001,\sub{1}}) = W(\Phi_{0101,\sub{1}}) = W(\Phi_{1010,\sub{1}})$;
\item
$W(\sub{3}) = W(\Phi_{0011,\sub{3}}) = W(\Phi_{1100,\sub{3}}) = W(\Phi_{0110,\sub{3}}) = W(\Phi_{1001,\sub{3}})$.
\end{itemize}
Consequently, we have
$\left\{\begin{smallmatrix}
a' = W(\subinmatrix{2}) + W(\subinmatrix{3}) \\
b' = W(\subinmatrix{1}) + W(\subinmatrix{3}) \\
c' = W(\subinmatrix{1}) + W(\subinmatrix{2})
\end{smallmatrix}\right.$. $W(\sub{2})$, $W(\sub{1})$, and $W(\sub{3})$ are all nonnegative due to the fact that the weight of every directed Eulerian partition has a nonnegative weight. Therefore, $a' \le b' + c'$, $b' \le a' + c'$, and $c' \le a' + b'$. This is to say, $f \in \mathcal{F}_\le$.
\end{proof}
\begin{proof}[Proof of \thmref{conf_2}]
Inheriting the notations from the above proof, we have $w(\sub{3}) = 0$ when $c = a + b$ for each vertex, which is to say no \enquote{crossing} can be made at any vertex in any Eulerian partition.
Due to planarity, a walk $e_1 \leadsto e_3$ must cross a walk $e_2 \leadsto e_4$ at a vertex, thus $W(\sub{3}) = W(\Phi_{0011,\sub{3}}) = 0$. Therefore, $c' = a' + b'$.
\end{proof}

\begin{rem}
\thmref{conf_1} not only serves as a depiction of the divergence of six-vertex models under different parameters separated by the phase transition threshold, but also helps us in \secref{sec:fpras} to bound the mixing time of a Markov chain so that approximately counting via sampling \cite{JERRUM1986169} leads to an FPRAS.
\end{rem}

%% file: fpras.tex
\documentclass[paper]{subfiles}

In this section we prove the following theorem.
\begin{thm}\label{fpras}
There is an FPRAS for computing $\Holant\left(\neq_2 | \mathcal{F}_{\le^2}\right)$.
\end{thm}

For simplicity we prove Theorem~\ref{fpras} only for the case
where all signatures of arity 4 used in the right-hand side
are from a fixed finite subset $\mathcal{F} \subset
 \mathcal{F}_{\le^2}$, i.e.,   we show that there is an 
FPRAS for computing $\Holant\left(\neq_2 | \mathcal{F}\right)$.
With some care the more general statement in Theorem~\ref{fpras} can also be
proved. 

We use the common approach to approximate counting via almost uniform sampling~\cite{JERRUM1986169} using a rapidly mixing Markov chain~\cite{doi:10.1137/0218077, Dyer:1991:RPA:102782.102783, Sinclair92improvedbounds, Jerrum-book}.

Our Markov chain $\mathcal{MC}$ is described in the setting of $\Holant\left(\neq_2 | \mathcal{F}_{\le^2}\right)$.
Let  $G = (V,U,E)$ be the underlying bipartite graph of
an instance of $\Holant\left(\neq_2 | \mathcal{F}_{\le^2}\right)$.
For simplicity we prove Theorem~\ref{fpras}
Each vertex in $V$ is assigned  $(\neq_2)$;
each vertex  $u \in U$ is assigned a signature 
$f_u \in \mathcal{F}_{\le^2}$.  An \textit{assignment} $\sigma$ assigns a
value in $\{0, 1\}$ to each edge $e \in E$.
The  state space of $\mathcal{MC}$ is $\Omega = \Omega_0 \cup \Omega_2$, which 
consists of  \enquote{perfect}  or \enquote{near-perfect} assignments
to $E$: All assignments satisfy the \enquote{two-0 two-1}
  ice rule  at every vertex $u \in U$ of degree 4.
We also  insist that all assignments satisfy the  \enquote{one-0 one-1}
at every $v \in V$ with \emph{possibly exactly two} exceptions.
Assignments in $\Omega_0$ have no exceptions, and are \enquote{perfect}.
Assignments in $\Omega_2$ have exactly two exceptions, and are
\enquote{near-perfect}.
Thus any $\sigma \in \Omega_0$ sastifies all $(\neq_2)$ on $V$,
and  any $\sigma \in \Omega_2$ sastifies all $(\neq_2)$ on $V -\{v', v''\}$
for some two vertices $v', v'' \in V$ where it satisfies $(=_2)$
(which outputs 1 on inputs 00, 11 and outputs 0 on 01, 10).

For any assignment $\sigma
\in \Omega$ and any subset $S \subseteq \Omega$,
define the \textit{weight} function $\mathcal{W}$
by 
 $\mathcal{W}(\sigma) = \prod_{u \in U} f_u(\sigma |_{E(v)})$ and $\mathcal{Z}(S) = \sum_{\sigma \in S} \mathcal{W}(\sigma)$.
Then the \textit{Gibbs measure} for $\Omega$ is defined
by $\pi(\sigma) = \frac{\mathcal{W}(\sigma)}{\mathcal{Z}(\Omega)}$,
assuming $\mathcal{Z}(\Omega)>0$.
Observe that if a state $\sigma \in \Omega_2$ 
assigns 00 to both edges incident to $v' \in V$ (satisfying $(=_2)$ at $v'$)
 then it must assign 11 to  both edges incident to $v'' \in V$, and vice versa. Indeed, having 00 at $v'$ models the fact that $v'$ has two arrows going out (to degree-4 vertices in $U$). To maintain the property that
 the number of incoming arrows is equal to the number of outgoing arrows everywhere else, $v''$ must have two arrows coming in, which is equivalent to having 11  at $v''$ in the Holant setting. 
An example state is shown in \figref{fig:congestion_state}.

Transitions in $\mathcal{MC}$ are comprised of three types of moves.
Suppose $\sigma \in \Omega_0$.
An  $\Omega_0$-to-$\Omega_2$ move from $\sigma$
 takes a degree 4 vertex $u \in U$ and
two incident edges $e' = (v', u), e'' = (v'', u) \in V \times U$
satisfying $\{\sigma(e'), \sigma(e'')\}  = \{0, 1\}$,
and changes it to $\sigma_2 \in \Omega_2$
which flips both $\sigma(e')$ and $\sigma(e'')$.
The effect is that we still have $\{\sigma_2(e'), \sigma_2(e'')\}  = \{0, 1\}$,
but at $v'$ and $v''$, $\sigma_2$ satisfies $(=_2)$ instead.
An $\Omega_2$-to-$\Omega_0$ move is the opposite.
An $\Omega_2$-to-$\Omega_2$ move is, intuitively,  to 
\textit{shift}  one $(=_2)$ from one vertex  $v' \in V$ 
to another $v^* \in V$, where for some $u \in U$, 
 $v'$ and $v^*$ are both incident to $u$
and  the \enquote{two-0 two-1} rule at $u$ is preserved. 
 Formally, let  $\sigma \in \Omega_2$ be a near-perfect assignment
with $v', v'' \in V$ being the two exceptional vertices (i.e.,
$\sigma$ satisfies $(=_2)$ at $v'$ and $v''$). 
Let $v^* \in V - \{v', v''\}$ be such that
for some $u \in U$, both $e' = (v', u), e^* = (v^*, u) \in E$,
and $\{\sigma(e'), \sigma(e^*)\}  = \{0, 1\}$.
Then an $\Omega_2$-to-$\Omega_2$ move changes $\sigma$ to $\sigma^*$
by flipping  both $\sigma(e')$ and $\sigma(e^*)$.
The effect is that we still have $\{\sigma^*(e'), \sigma^*(e^*)\}  = \{0, 1\}$,
but  $\sigma^*$ satisfies $(\neq_2)$ at $v'$ 
and  $(=_2)$  at $v^*$.  Note that $\sigma^*$ continues to satisfy
 $(=_2)$ at $v''$.

The above describes a symmetric binary relation \textit{neighbor} ($\sim$)
 on $\Omega$.  No two states in $\Omega_0$ are neighbors.
Set $n = |U|$.
The transition probabilities $P(\cdot, \cdot)$ of $\mathcal{MC}$ are \textit{Metropolis} moves between neighbouring states:
\[
P(\sigma_1, \sigma_2) = 
\left\{
\begin{array}{ll}
\frac{1}{n^2}\min\left(1, \frac{\pi(\sigma_2)}{\pi(\sigma_1)}\right) & \text{if }\sigma_2 \sim \sigma_1; \\
1 - \frac{1}{n^2}\sum_{\sigma' \sim \sigma_1}\min\left(1, \frac{\pi(\sigma')}{\pi(\sigma_1)}\right) & \text{if } \sigma_1 = \sigma_2; \\
0 & \text{otherwise.}
\end{array}
\right.
\]
$\mathcal{MC}$ is aperiodic due to the \enquote{lazy} movement; one can verify that $\mathcal{MC}$ is irreducible by creating, shifting, and merging of a pair of $(=_2)$'s; as the transitions are Metropolis moves, detailed balance conditions are satisfied with regard to $\pi$. 
By results from~\cite{doi:10.1137/0218077, Sinclair92improvedbounds},
such a Markov chain is rapidly mixing if there is a \textit{flow} whose congestion can be bounded by a polynomial in $n$.
\begin{lem}\label{congestion}
Assume $\mathcal{Z}(\Omega_0) > 0$. There is a flow on $\Omega$ with congestion at most $O\left(n^3\left(\frac{\mathcal{Z}(\Omega)}{\mathcal{Z}(\Omega_0)}\right)^2\right)$, using paths of length $O(n)$.
\end{lem}
\begin{proof}
The idea is to design a flow $\mathfrak{F}: \mathcal{P} \rightarrow \mathbb{R}^+$ from $\Omega_2$ to $\Omega_0$ which satisfies
\[\sum_{p \in \mathcal{P}_{\sigma_2\sigma_0}} \mathfrak{F}(p) = \pi(\sigma_2)\pi(\sigma_0)\text{, \; for all } \sigma_2 \in \Omega_2, \sigma_0 \in \Omega_0,\]
where $\mathcal{P}_{\sigma_2\sigma_0}$ is defined to be a set of simple directed paths from $\sigma_2$ to $\sigma_0$ in $\mathcal{MC}$
and $\mathcal{P} = \bigcup_{\sigma_2 \in \Omega_2, \sigma_0 \in \Omega_0}\mathcal{P}_{\sigma_2\sigma_0}$. 
Once the congestion of $\mathfrak{F}$ from $\Omega_2$ to $\Omega_0$ is polynomially bounded, so is the flow from $\Omega_0$ to $\Omega_2$ by symmetric construction. Moreover, there is a flow from $\Omega_2$ to $\Omega_2$ (or from $\Omega_0$ to $\Omega_0$) whose congestion can also be polynomially bounded by randomly picking an intermediate state in $\Omega_0$ (or $\Omega_2$, respectively). Thus we have a flow on $\Omega$ with polynomially bounded congestion. This technique has been used in \cite{Jerrum:2004:PAA:1008731.1008738, DBLP:journals/corr/abs-1301-2880}.
In the following we show that the congestion of $\mathfrak{F}$
from  $\Omega_2$ to $\Omega_0$ is
 bounded by $O(n^3)\frac{\mathcal{Z}(\Omega_2)}{\mathcal{Z}(\Omega_0)}$.
Then the bound in the lemma for a flow on $\Omega$  follows. 

To describe the flow $\mathfrak{F}$, we first specify the
sets of paths that are going to take the flow.
In line with the definition of $\Omega_0$ and $\Omega_2$, 
we define $\Omega_4$ to be the set of assignments where there are exactly four violations of $(\neq_2)$ in $V$. Let $\Omega' = \Omega_0 \cup \Omega_2 \cup \Omega_4$.
For $\sigma, \sigma' \in \Omega'$, let  $\sigma \oplus \sigma'$
denote the  \textit{symmetric difference} (or bitwise  XOR),
where we view $\sigma$ and $\sigma'$ as two bit strings in $\{0, 1\}^{|E|}$.
This is a 0-1 assignment to the edge set of the bipartite graph
$G = (V, U, E)$. We also treat $\sigma \oplus \sigma'$ as an edge subset
of $E$ (corresponding to bit 
positions having bit 1, where  $\sigma$ and $\sigma'$
assign opposite values), and this defines an induced
subgraph of $G$. Since at every $u \in U$ of degree 4, the 
\enquote{two-0 two-1} rule is satisfied by both  $\sigma$ and $\sigma'$,
this induced subgraph has even degree (0, 2, or 4) at every $u \in U$.

Denote by $U_4 \subseteq U$ the degree-4 vertices in $\sigma \oplus \sigma'$.
Then there are exactly $2^{|U_4|}$ Eulerian partitions for $\sigma \oplus \sigma'$.
Recall that an Eulerian partition of $\sigma \oplus \sigma'$ is uniquely determined by a family of pairings on $U_4$.
This is a 1-1 correspondence and we will identify the two sets.
For any  pairing in $\{\sub{1}, \sub{2}, \sub{3}\}$ on a vertex
$u$ with signature matrix $M(f_u) = \left[\begin{smallmatrix} & & & a \\ & b & c & \\ & c & b & \\ a & & &\end{smallmatrix}\right]$, define
the weight function $\mathfrak{w}$ for pairings as follows,
$\left\{\begin{smallmatrix}
\mathfrak{w}(\subinmatrix{1}) = \frac{-a^2 + b^2 + c^2}{2} \\
\mathfrak{w}(\subinmatrix{2}) = \frac{a^2 - b^2 + c^2}{2} \\
\mathfrak{w}(\subinmatrix{3}) = \frac{a^2 + b^2 - c^2}{2}
\end{smallmatrix}\right.$,
or equivalently
$\left\{\begin{smallmatrix}
a^2 = \mathfrak{w}(\subinmatrix{2}) + \mathfrak{w}(\subinmatrix{3}) \\
b^2 = \mathfrak{w}(\subinmatrix{1}) + \mathfrak{w}(\subinmatrix{3}) \\
c^2 = \mathfrak{w}(\subinmatrix{1}) + \mathfrak{w}(\subinmatrix{2})
\end{smallmatrix}\right.$.
Note that when $f_u \in \mathcal{F}_{\le^2}$, $\mathfrak{w}$ takes \textit{nonnegative} values.
Let $\Phi_{\sigma \oplus \sigma'}$ be the set of Eulerian partitions for $\sigma \oplus \sigma'$.
For $\varphi \in \Phi_{\sigma \oplus \sigma'}$, define
\[\mathfrak{W}(\sigma, \sigma', \varphi) := \left(\prod_{u \in U\setminus U_4}f_u\left(\sigma |_{E(u)}\right)f_u\left(\sigma' |_{E(u)}\right)\right)\left(\prod_{u \in U_4}\mathfrak{w}(\varphi(u))\right).\]
Then for all distinct $\sigma, \sigma' \in \Omega'$, we have
\begin{align*}
\sum_{\varphi \in \Phi_{\sigma \oplus \sigma'}}\mathfrak{W}(\sigma, \sigma', \varphi)
& = \sum_{\varphi \in \Phi_{\sigma \oplus \sigma'}} \left(\prod_{u \in U\setminus U_4}f_u\left(\sigma |_{E(u)}\right)f_u\left(\sigma' |_{E(u)}\right)\right)\left(\prod_{u \in U_4}\mathfrak{w}(\varphi(u))\right) \\
& = \left(\prod_{u \in U\setminus U_4}f_u\left(\sigma |_{E(u)}\right)f_u\left(\sigma' |_{E(u)}\right)\right) \left(\sum_{\varphi \in \Phi_{\sigma \oplus \sigma'}} \prod_{u \in U_4}\mathfrak{w}(\varphi(u)) \right) \\
& = \left(\prod_{u \in U\setminus U_4}f_u\left(\sigma |_{E(u)}\right)f_u\left(\sigma' |_{E(u)}\right)\right) \left(\prod_{u \in U_4}f_u\left(\sigma |_{E(u)}\right)f_u\left(\sigma' |_{E(u)}\right)\right) \\
& = \prod_{u \in U} f_u\left(\sigma |_{E(u)}\right)f_u\left(\sigma' |_{E(u)}\right) \\
& = \mathcal{W}(\sigma)\mathcal{W}(\sigma').
\end{align*}
The equality from line 2 to line 3 is due to
the following: when the degree (in the
induced subgraph  $\sigma \oplus \sigma'$)  of a vertex $u \in U$ is 4, $\sigma$ and $\sigma'$ must take the same value at $u$,
since one represents a total reversal of all arrows of another; 
thus $f_u\left(\sigma |_{E(u)}\right)f_u\left(\sigma' |_{E(u)}\right)$ is in $\{a^2, b^2, c^2\}$. Then
\[\prod_{u \in U_4}
f_u\left(\sigma |_{E(u)}\right)f_u\left(\sigma' |_{E(u)}\right)
=  \sum_{\varphi \in \Phi_{\sigma \oplus \sigma'}} \prod_{u \in U_4}\mathfrak{w}(\varphi(u))\]
 is obtained by using the sum expressions for $a^2, b^2$ and
$c^2$ in terms of 
$\mathfrak{w}(\sub{1}),
\mathfrak{w}(\sub{2})$, and
$\mathfrak{w}(\sub{3})$,
 and then expressing the product-of-sums as a sum-of-products.

Now we are ready to specify the \enquote{paths} which take nonzero flow from $\sigma_2 \in \Omega_2$ to $\sigma_0 \in \Omega_0$.
In order to transit from $\sigma_2$ to $\sigma_0$, paths in $\mathcal{P}_{\sigma_2\sigma_0}$ go through states in $\Omega$ that gradually decrease the number of conflicting assignments along walks and circuits in $\sigma_2 \oplus \sigma_0$.
We first specify a total order on $E$, the set of edges of $G$.
This induces a total order on circuits by lexicographic order. 
In the induced subgraph $\sigma_2 \oplus \sigma_0$, exactly
 two vertices in $V$ have degree 1 (called \textit{endpoints}) and
all other vertices have degree 2 or degree 4.
The set of paths in $\mathcal{P}_{\sigma_2\sigma_0}$ are  designed to be in 1-to-1 correspondence with elements in $\Phi_{\sigma_2 \oplus \sigma_0}$.
Given any family of pairings $\varphi \in \Phi_{\sigma_2 \oplus \sigma_0}$,
we have a unique decomposition of the induced 
subgraph $\sigma_2 \oplus \sigma_0$ as an edge disjoint union
of one walk 
$[e_1] (v_1, e'_1, u_1, e_2, v_2, e'_2, u_2, \ldots, e_k, v_k)  [e'_k]$
(where $e_1$ and $e'_k$ are not part of the walk),
 and zero or more edge disjoint circuits, which are
ordered lexicographically.
Here $v_i \in V$ and $u_i \in U$, and we may assume
$\sigma_2(e_1) = \sigma_2(e'_1) =0$, $\sigma_2(e_2)=1, \sigma_2(e'_2)=0,
\ldots,  \sigma_2(e_k)=\sigma_2(e'_k)=1$.
So the two exceptional vertices are $v_1$ and $v_k$, where
$\sigma_2$ satisfies $(=_2)$.
The unique path $p_\varphi$ first \enquote{pushes}  the $(=_2)$
from $v_1$, to $v_2$, then to $v_3, \ldots, v_{k-1}$, and then \enquote{merge} 
at $v_{k}$, arriving at a configuration in $\Omega_0$.
Then  $p_\varphi$  reverses all arrows on each
circuit in lexicographic order, and within each circuit $C$
it starts at the  least edge $e$ (according to the edge order)
and reverses all arrows
on $C$
in the direction defined by the starting cyclic orientation of $\sigma_2$. 
(Technically it flips a pair of incident edges to vertices in $U$
in each step.)
%
Such paths  $p_\varphi$  are well-defined and
are valid paths in $\mathcal{MC}$ since along any path every state is in $\Omega = \Omega_0 \cup \Omega_2$ and every move is a valid transition defined in $\mathcal{MC}$.
With regard to the flow distribution, the  flow value
put on $p_\varphi$ is $\frac{\mathfrak{W}(\sigma_2, \sigma_0, \varphi)}{\left(\mathcal{Z}(\Omega)\right)^2}$, making
the following hold for all $\sigma_2 \in \Omega_2, \sigma_0 \in \Omega_0$:
\begin{align*}
\sum_{p_\varphi \in \mathcal{P}_{\sigma_2\sigma_0}} \mathfrak{F}(p_\varphi) & = \sum_{\varphi \in \Phi_{\sigma_2 \oplus \sigma_0}}\frac{\mathfrak{W}(\sigma_2, \sigma_0, \varphi)}{\left(\mathcal{Z}(\Omega)\right)^2} \\
& = \frac{\mathcal{W}(\sigma_2)\mathcal{W}(\sigma_0)}{\left(\mathcal{Z}(\Omega)\right)^2} \\
& = \pi(\sigma_2)\pi(\sigma_0)
\end{align*}

For any transition $(\sigma', \sigma'')$ where $\sigma' \neq \sigma''$, we have $P(\sigma', \sigma'') = \frac{1}{n^2}\min\left(1, \frac{\pi(\sigma'')}{\pi(\sigma')}\right) = \Omega\left(\frac{1}{n^2}\right)$, as $\frac{\pi(\sigma'')}{\pi(\sigma')}$ is a constant. (This is a constant because we have restricted
the signatures $f_u$ to be from a fixed finite set $\mathcal{F}$.)
 Let $H_{\sigma'} = \{\sigma_2 \oplus \sigma_0 \ |\ \sigma_2 \in \Omega_2, \sigma_0 \in \Omega_0, \exists \varphi \in \Phi_{\sigma_2 \oplus \sigma_0} \text{ s.t. } \sigma' \in p_\varphi\}$.
The congestion of $\mathfrak{F}$ is
\begin{align*}
& \max_{\text{transition }(\sigma', \sigma'')}\frac{1}{\pi(\sigma')P(\sigma', \sigma'')}
\sum_{\substack{\sigma_2 \in \Omega_2 \\ \sigma_0 \in \Omega_0}}
\sum_{\substack{p_\varphi \in \mathcal{P}_{\sigma_2 \sigma_0} \\ p_\varphi \ni (\sigma',\sigma'')}}\frac{\mathfrak{W}(\sigma_2, \sigma_0, \varphi)}{\left(\mathcal{Z}(\Omega)\right)^2} \\
\le & \max_{\sigma' \in \Omega}\frac{O(n^2)}{\mathcal{W}(\sigma') \mathcal{Z}(\Omega)}
\sum_{\substack{\sigma_2 \in \Omega_2 \\ \sigma_0 \in \Omega_0}}
\sum_{\substack{\varphi \in \Phi_{\sigma_2 \oplus \sigma_0} \\ p_\varphi \ni \sigma'}}
\mathfrak{W}(\sigma_2, \sigma_0, \varphi) \\
\le & \max_{\sigma' \in \Omega}\frac{O(n^2)}{\mathcal{W}(\sigma') \mathcal{Z}(\Omega)}
\sum_{\sigma_2 \in \Omega_2}\sum_{\eta \in H_{\sigma'}}
\sum_{\varphi \in \Phi_\eta}
\mathfrak{W}(\sigma_2, \sigma_2 \oplus \eta, \varphi).
\end{align*}
Fix any $\sigma' \in \Omega$.
For any $\sigma_2 \in \Omega_2$, and $\eta \in H_{\sigma'}$ 
consisting of exactly one connected component with two endpoints of degree 1
and all other vertices having even degree
(and zero or more connected components of even degree vertices), observe that $\sigma' \oplus \eta \in \Omega'$. Indeed, if $\sigma' \in \Omega_0$ then $\sigma' \oplus \eta \in \Omega_2$; if $\sigma' \in \Omega_2$ then 
depending on whether $\sigma'$ 
\setlist[enumerate]{itemsep=-1mm}
\begin{enumerate}[(1)]
\item is $\sigma_2$, or
\item appears in the process of reversing arrows on the walk with two endpoints, or
\item appears after reversing arrows on the walk with endpoints,
\end{enumerate}
$\sigma' \oplus \eta$ lies in $\Omega_0$, $\Omega_2$, or $\Omega_4$, respectively.
For the edges not in $\eta$, $\sigma'$ agrees with $\sigma_2$ and $\sigma_2 \oplus \eta$ as the path $p_\varphi$ never \enquote{touches} them, and so does $\sigma' \oplus \eta$.
Recall that
\[\mathfrak{W}(\sigma_2, \sigma_2 \oplus \eta, \varphi) = \left(\prod_{u \in U\setminus U_4}f_u\left(\sigma_2 |_{E(u)}\right)f_u\left( (\sigma_2 \oplus \eta) |_{E(u)}\right)\right)\left(\prod_{u \in U_4}\mathfrak{w}(\varphi(u))\right).\] 
For every degree-0 vertex $u \in U$ (this notion of degree is 
in terms of the induced subgraph $\eta$, thus a degree-0 vertex $u \in U$ 
is not in the induced subgraph $\eta$), $f_u$ takes the same value in all $\sigma_2$, $\sigma_2 \oplus \eta$, $\sigma'$, and $\sigma' \oplus \eta$.
For every degree-2 vertex $u \in U$, assuming $M(f_u) = \left[\begin{smallmatrix} & & & a \\ & b & c & \\ & c & b & \\ a & & &\end{smallmatrix}\right]$,
$f_u\left(\sigma_2 |_{E(u)}\right)$ and $f_u\left( (\sigma_2 \oplus \eta) |_{E(u)}\right)$ take two different elements in $\{a, b, c\}$.  Meanwhile, $f_u\left(\sigma' |_{E(u)}\right)$ and $f_u\left(\sigma' \oplus \eta |_{E(u)}\right)$ also take these two elements (possibly in the opposite order).
For example, in~\figref{fig:xor_degree2} the 
two solid edges are in $\eta$ and assignments on the two dotted edges are 
shared by $\sigma_2$ and $\sigma_2 \oplus \eta$, as well as
$\sigma'$ and $\sigma' \oplus \eta$.
 On the two solid edges
$\sigma'$ either agrees with  $\sigma_2$  or  $\sigma_2 \oplus \eta$,
and $\sigma' \oplus \eta$ is its reversal and agrees with the other.
For every degree-4 vertex $u \in U$, $\mathfrak{w}(\varphi(u))$ takes the same value in $\mathfrak{W}(\sigma_2, \sigma_2 \oplus \eta, \varphi)$ and $\mathfrak{W}(\sigma', \sigma' \oplus \eta, \varphi)$ as the weight only 
depends on $\varphi(u)$, the pairing at $u$.

\captionsetup[subfigure]{labelformat=parens}
\begin{figure}[h!]
\centering
\begin{subfigure}[b]{0.15\linewidth}
\centering\includegraphics[width=\linewidth]{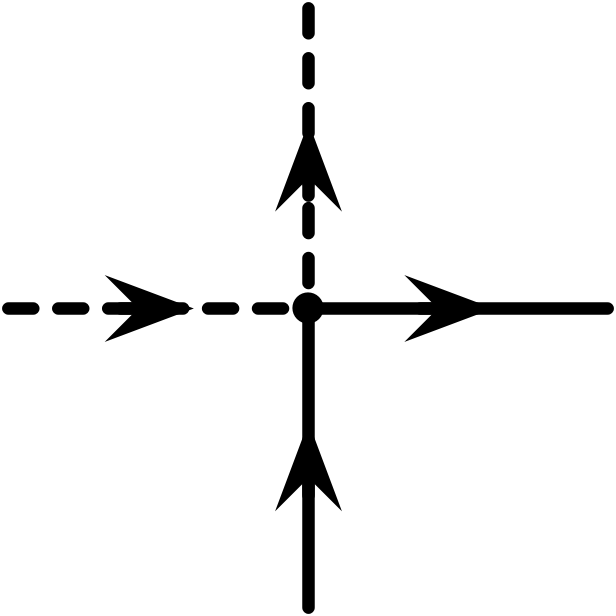}\caption{$\sigma_2$.}\label{fig:xor_degree2_before}
\end{subfigure}
\hspace{0.2\linewidth}
\begin{subfigure}[b]{0.15\linewidth}
\centering\includegraphics[width=\linewidth]{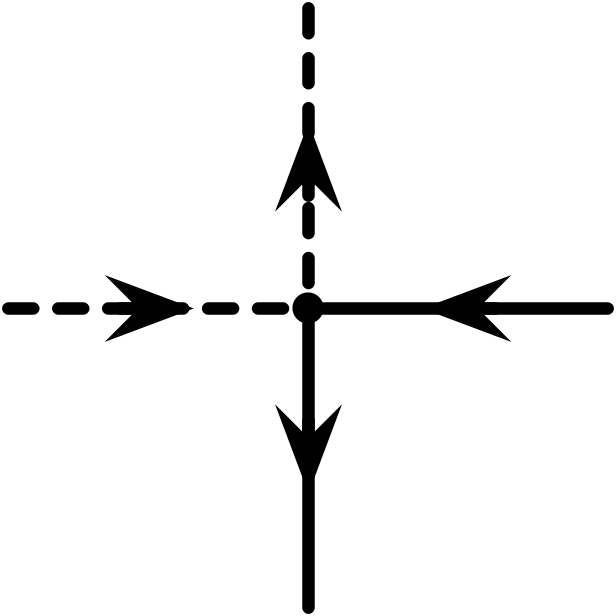}\caption{$\sigma_2 \oplus \eta$.}\label{fig:xor_degree2_after}
\end{subfigure}
\caption{}\label{fig:xor_degree2}
\end{figure}

By the above argument, we established that $\mathfrak{W}(\sigma_2, \sigma_2 \oplus \eta, \varphi) = \mathfrak{W}(\sigma', \sigma' \oplus \eta, \varphi)$.
Therefore, the congestion of $\mathfrak{F}$ can be bounded by
\begin{align*}
& \max_{\sigma' \in \Omega}\frac{O(n^2)}{\mathcal{W}(\sigma') \mathcal{Z}(\Omega)}
\sum_{\sigma_2 \in \Omega_2}\sum_{\eta \in H_{\sigma'}}
\sum_{\varphi \in \Phi_\eta}
\mathfrak{W}(\sigma', \sigma' \oplus \eta, \varphi) \\
\le & \max_{\sigma' \in \Omega}\frac{O(n^2) |E|}{\mathcal{W}(\sigma') \mathcal{Z}(\Omega)}
\sum_{\eta \in H_{\sigma'}}
\sum_{\varphi \in \Phi_\eta}
\mathfrak{W}(\sigma', \sigma' \oplus \eta, \varphi) \\
\le & \max_{\sigma' \in \Omega}\frac{O(n^3)}{\mathcal{W}(\sigma') \mathcal{Z}(\Omega)}
\sum_{\eta \in H_{\sigma'}}
\mathcal{W}(\sigma')\mathcal{W}(\sigma' \oplus \eta) \\
= & \max_{\sigma' \in \Omega}\frac{O(n^3)}{\mathcal{Z}(\Omega)}
\sum_{\eta \in H_{\sigma'}}\mathcal{W}(\sigma' \oplus \eta) \\
\le & \max_{\sigma' \in \Omega}\frac{O(n^3)}{\mathcal{Z}(\Omega)}
\sum_{\sigma \in \Omega'}\mathcal{W}(\sigma) \\
= & \max_{\sigma' \in \Omega}O(n^3)\frac{\mathcal{Z}(\Omega')}{\mathcal{Z}(\Omega)}.
\end{align*}
By a standard argument as in \cite{doi:10.1137/0218077, Mihail1996, DBLP:journals/corr/abs-1301-2880}, $\frac{\mathcal{Z}(\Omega_4)}{\mathcal{Z}(\Omega_2)} \le \frac{\mathcal{Z}(\Omega_2)}{\mathcal{Z}(\Omega_0)}$. Therefore, the congestion is bounded by $O(n^3)\frac{\mathcal{Z}(\Omega_2)}{\mathcal{Z}(\Omega_0)}$.
Note that in each path, no edge is flipped more than once, so the length is $O(n)$.  
\end{proof}

In order to show $\mathcal{MC}$ is rapidly mixing, we 
need to show $\frac{\mathcal{Z}(\Omega_2)}{\mathcal{Z}(\Omega_0)}$ is polynomially bounded. This bound is also needed to get
an FPRAS from a rapidly mixing Markov chain
in $\Omega$, since ultimately we are only interested in $\Omega_0$. Such a
bound is a corollary of \thmref{conf_1}.

\captionsetup[subfigure]{labelformat=parens}
\begin{figure}[h!]
\centering
\begin{subfigure}[b]{0.4\linewidth}
\centering\includegraphics[width=\linewidth]{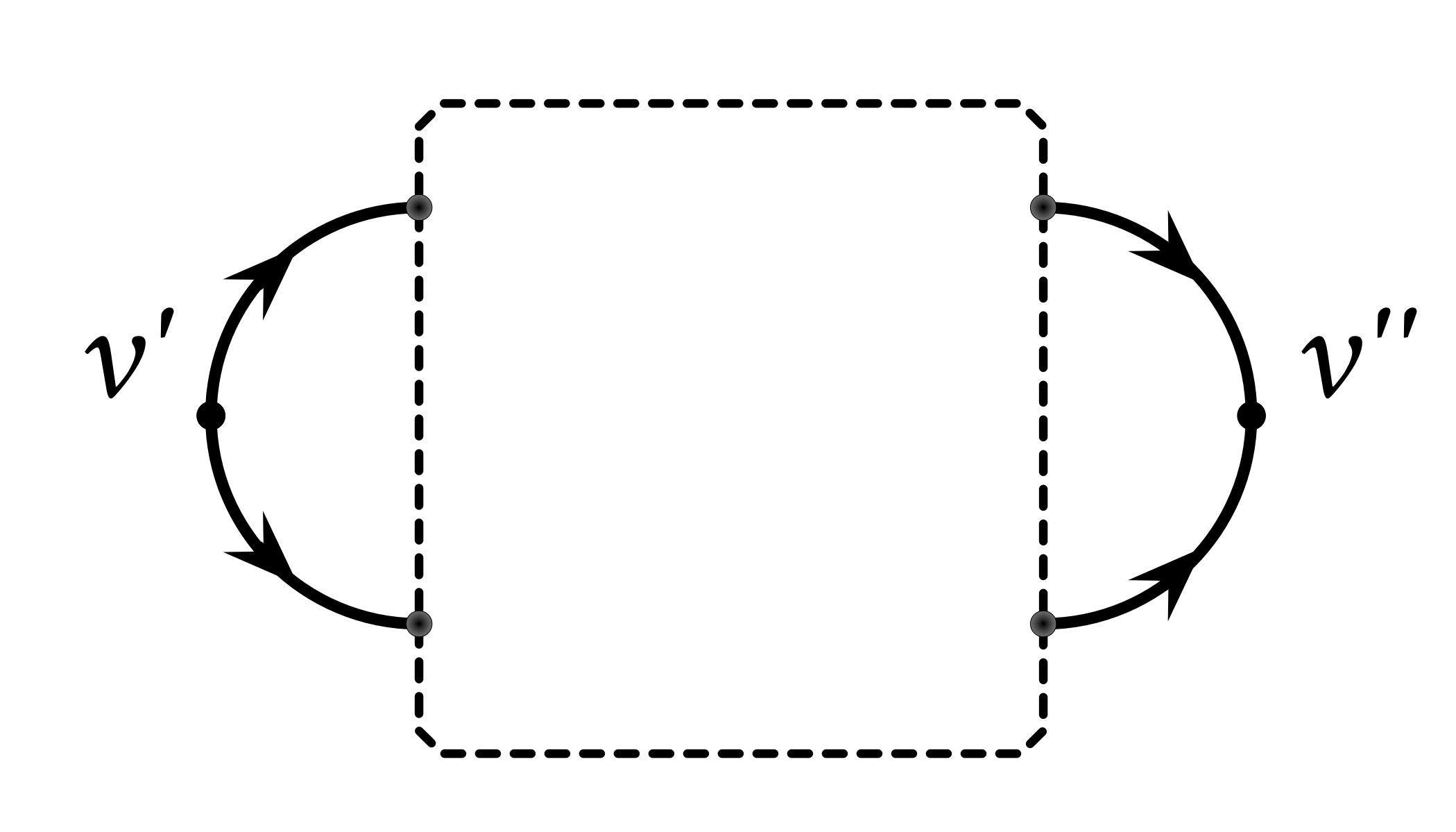}\caption{A state in $\Omega_2$ with $(=_2)$'s at $v'$ and $v''$.}\label{fig:congestion_state}
\end{subfigure}
\hspace{0.05\linewidth}
\begin{subfigure}[b]{0.4\linewidth}
\centering\includegraphics[width=\linewidth]{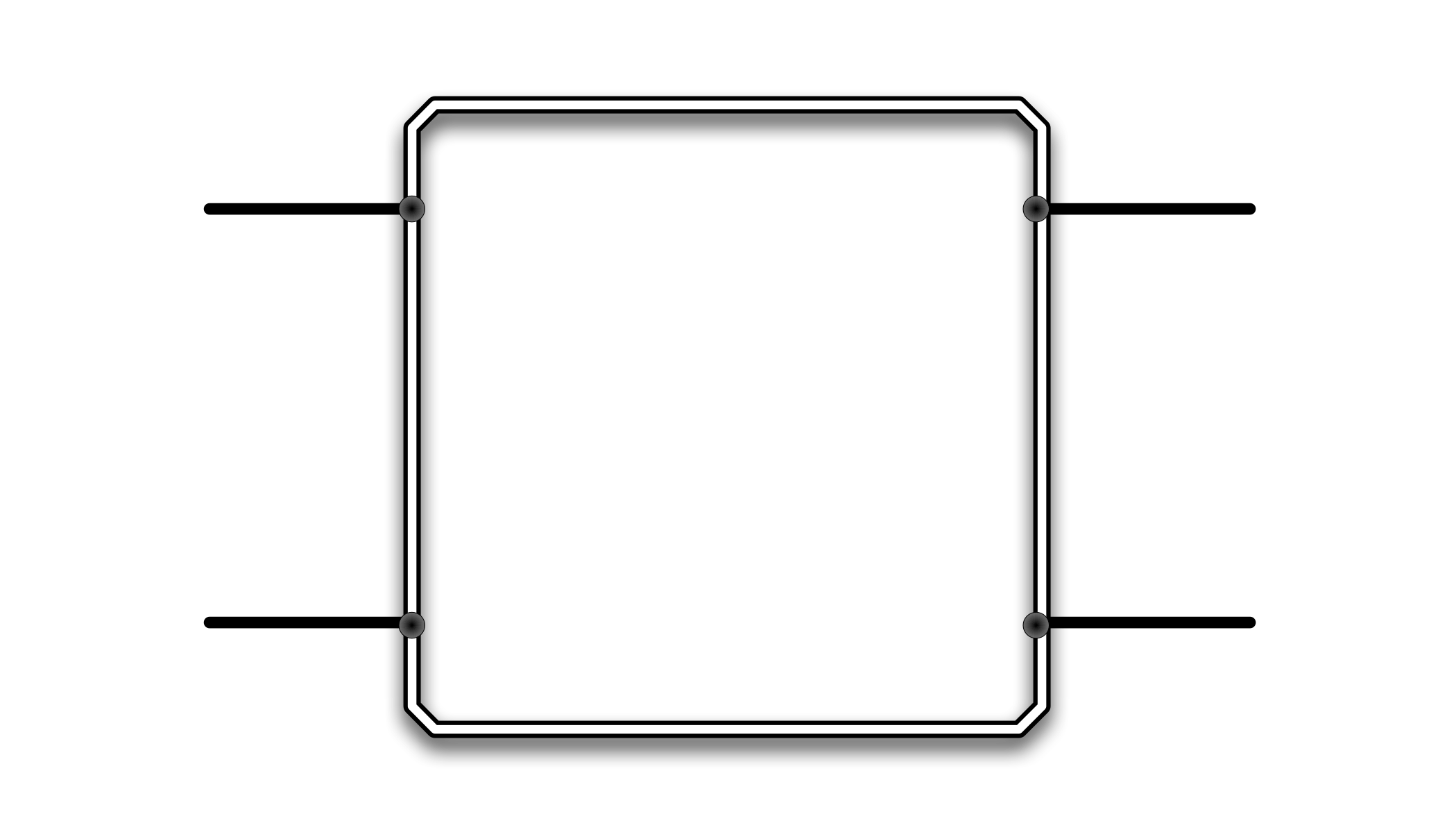}\caption{A gadget made by deleting $v'$ and $v''$.}\label{fig:congestion_gadget}
\end{subfigure}
\caption{}\label{fig:congestion}
\end{figure}

For each $\sigma \in \Omega_2$, there are exactly two vertices in $V$ 
satisfying $(=_2)$. Let $\Omega_2^{\{v', v''\}} \subseteq \Omega_2$ be 
the set of states in which $v', v''$ are these two vertices.
 We have $\frac{\mathcal{Z}(\Omega_2)}{\mathcal{Z}(\Omega_0)} = 
\sum_{\{v', v''\} \ \in \ {V \choose 2}}\frac{\mathcal{Z}(\Omega_2^{\{v', v''\}})}{\mathcal{Z}(\Omega_0)}$.
For any $\sigma \in \Omega_2^{\{v', v''\}}$, the local assignments around $v'$ and $v''$ must be 00 on one and 11 on the other. An example is in \figref{fig:congestion_state}. If we \enquote{delete} $v'$ and $v''$ as shown in \figref{fig:congestion_gadget}, we get a 4-ary gadget $g$ on the RHS of $\Holant\left(\neq_2 | \mathcal{F}_{\le^2}\right)$. Denote the signature matrix of $g$ by $M(g) = \left[\begin{smallmatrix} & & & a' \\ & b' & c' & \\ & c' & b' & \\ a' & & &\end{smallmatrix}\right]$, with the input order being counter-clockwise starting from the upper-left edge.
For this gadget $g$ we observe that: the states in $\Omega_2^{\{v', v''\}}$ where edges incident to $v'$ (also $v''$) take the same value contribute a total weight $(a' + a')$, i.e. $\mathcal{Z}(\Omega_2^{\{v', v''\}}) = 2a'$; the states in $\Omega_0$ where $v', v''$ satisfy $(\neq_2)$ have a total weight $\mathcal{Z}(\Omega_0) = 2b' + 2c'$.
Note that $\mathcal{F}_{\le^2} \subset \mathcal{F}_\le$.
By \thmref{conf_1} we know that for gadget $g$, $a' \le b' + c'$. Therefore, $\frac{\mathcal{Z}(\Omega_2^{\{v', v''\}})}{\mathcal{Z}(\Omega_0)} \le 1$. In total, $\frac{\mathcal{Z}(\Omega_2)}{\mathcal{Z}(\Omega_0)} \le \binom{|V|}{2}$. Thus we have the following corollary.
\begin{cor}\label{proportion}
$\frac{\mathcal{Z}(\Omega_2)}{\mathcal{Z}(\Omega_0)} = O(n^2)$.
\end{cor}

Combining \lemref{congestion} and \corref{proportion},
we conclude that $\mathcal{MC}$ is rapidly mixing, and $\Omega_0$, the set of valid six-vertex configurations, in 
total takes a non-negligible proportion in the stationary distribution.
As a consequence, we are able to efficiently sample six-vertex configurations according to the Gibbs measure on $\Omega_0$, and in the following algorithm we only work with states in $\Omega_0$.
We design the following algorithm to approximately compute $\Holant\left(\neq_2 | \mathcal{F}_{\le^2}\right)$ via sampling with the Markov chain $\mathcal{MC}$.
As we have argued in \secref{sec:theorems}, the partition function of six-vertex models can be viewed as the weighted sum of Eulerian partitions. For a vertex $v \in U$, the ratios among different pairings (\sub{1}, \sub{2}, and \sub{3}) in weighted Eulerian partitions can be uniquely determined by the ratios among different orientations (represented by $a$, $b$, and $c$) at $v$. As long as the partition function is not zero (this can be easily tested in P), there must be a pairing $\varrho$ showing up at $v$ with probability at least $\frac{1}{3}$ among all three pairings. Therefore, running $\mathcal{MC}$ on $G$, we can approximate, with a sufficient $1/{\rm poly(n)}$ precision, the probability of having $\varrho$ at $v$, denoted by $\Pr_v (\varrho)$.
Denote by $G_{v, \varrho}$ the signature grid with $v$ being split into $v_1$ and $v_2$, each assigned a $(\neq_2)$ and the edges reconnected according to $\varrho$. Write the partition function of $G_{v, \varrho}$ as $Z(G_{v, \varrho})$, we have $\Pr_v (\varrho) = w(\varrho)Z(G_{v, \varrho}) / Z(G)$ which means $Z(G) = w(\varrho)Z(G_{v, \varrho}) / \Pr_v (\varrho)$. To approximate $Z(G)$ it suffices to approximate $Z(G_{v, \varrho})$, which can be done by running $\mathcal{MC}$ on $G_{v, \varrho}$ and recursing.
Repeating this process for $|U|$ steps we decompose the graph $G$ into the base case, a set of disjoint cycles with even number of vertices, each assigned a $(\neq_2)$. The partition function of this cycle graph is just $2^C$ where $C$ is the number of cycles. By this self-reduction, the partition function for $G$ can be approximated.

Therefore, \thmref{fpras} is proved. Note that for the special case
$(1,1,1)$, the FPRAS by Mihail and Winkler is a reduction \cite{Mihail1996} to computing the number of perfect matchings in a bipartite graph. We 
give a direct algorithm using Markov chain Monte-Carlo.
\begin{rem}
There is an alternative derivation of a rapidly mixing Markov chain using the notion of \enquote{windability}~\cite{DBLP:journals/corr/abs-1301-2880, Huang:2016:CPM:2884435.2884473}, for the purpose of approximating $\Holant\left(\neq_2 | \mathcal{F}_{\le^2}\right)$. Readers are referred to the Appendix for a proof 
 that signatures in $\mathcal{F}_{\le^2}$ are \textit{windable}. The mixing rate of that Markov chain can be bounded using similar techniques introduced in this section.
\end{rem}

%% file: hardness.tex
\documentclass[paper]{subfiles}

\begin{thm}
If $f \in \mathcal{F}_>$,
then $\Holant\left(\neq_2 | f\right)$ does not have an FPRAS unless {\rm RP}
$=$ {\rm NP}.
\end{thm}
\begin{proof}
Let 3-MIS denote the NP-hard problem of computing the cardinality of a 
maximum independent set in a 3-regular graph \cite{GAREY1976237}.
We  reduce 3-MIS
 to approximating $\Holant\left(\neq_2 | f\right)$. 
Since $f \in \mathcal{F}_>$, all $a, b, c >0$.
Since the proof of NP-hardness for $\Holant\left(\neq_2 | f\right)$
is for general graphs (i.e., not necessarily planar),
we can permute the parameters so that $c > a + b$,
and normalize $c > b \ge a =1$.
Let $\gamma = \frac{c}{a+b}$. Then $\gamma > 1$.

Before proving this theorem we briefly state our idea.
Denote an instance of 3-MIS by $G = (V, E)$. For any independent set, no two 
adjacent vertices $u, v \in V$ can both appear. The only possible configurations for $u, v$ in any independent set $S$ are $(u \in S, v \not\in S)$, $(u \not\in S, v \in S)$, and $(u \not\in S, v \not\in S)$.
We want to encode this local constraint by a local  
  fragment of $G'$ in terms of configurations in the six-vertex model.
%

\captionsetup[subfigure]{labelformat=parens}
\begin{figure}[h!]
\centering
\begin{subfigure}[b]{0.32\linewidth}
\centering\includegraphics[width=0.75\linewidth]{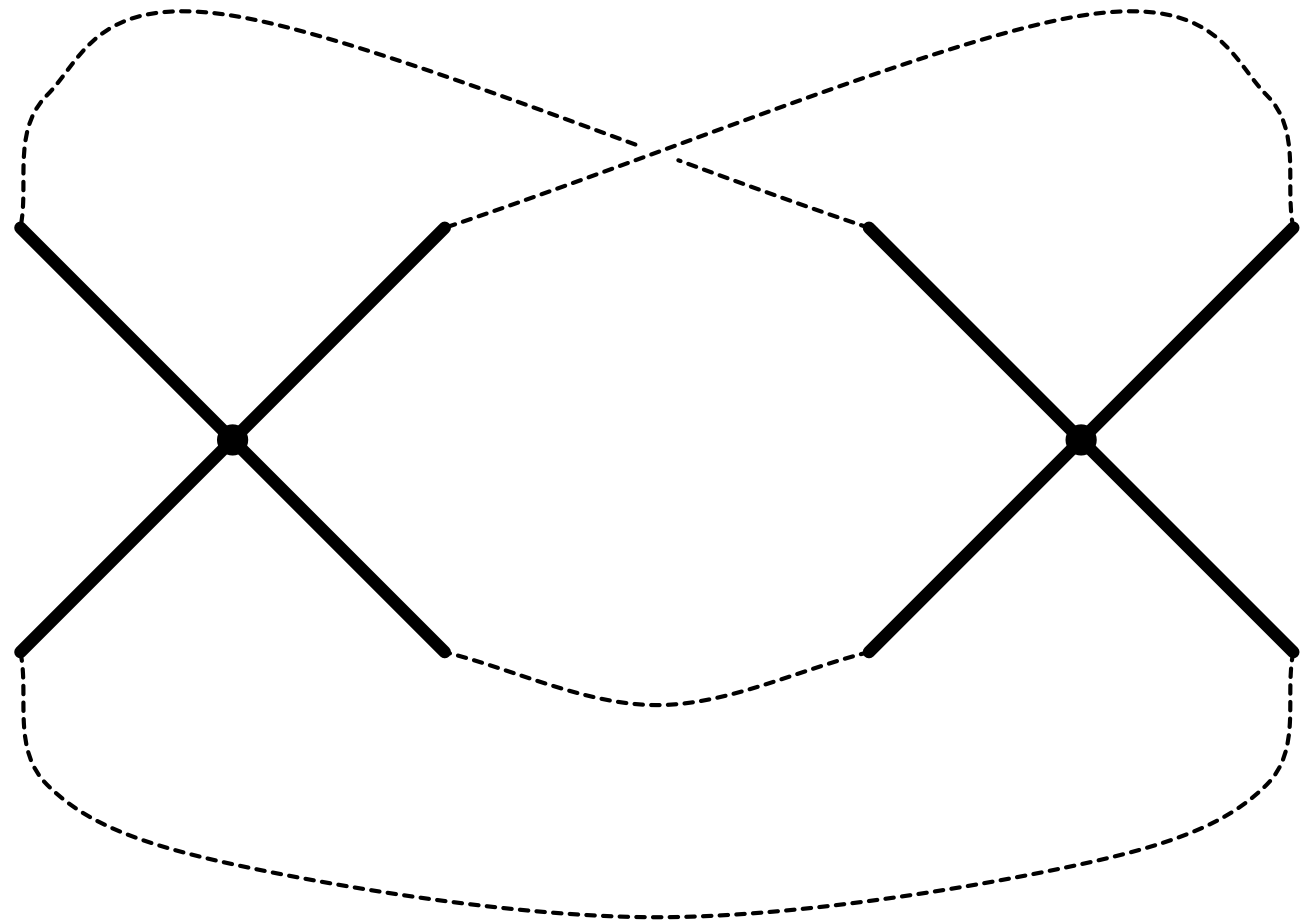}\caption{}\label{fig:hardness_single_undirected}
\end{subfigure}
\begin{subfigure}[b]{0.32\linewidth}
\centering\includegraphics[width=0.75\linewidth]{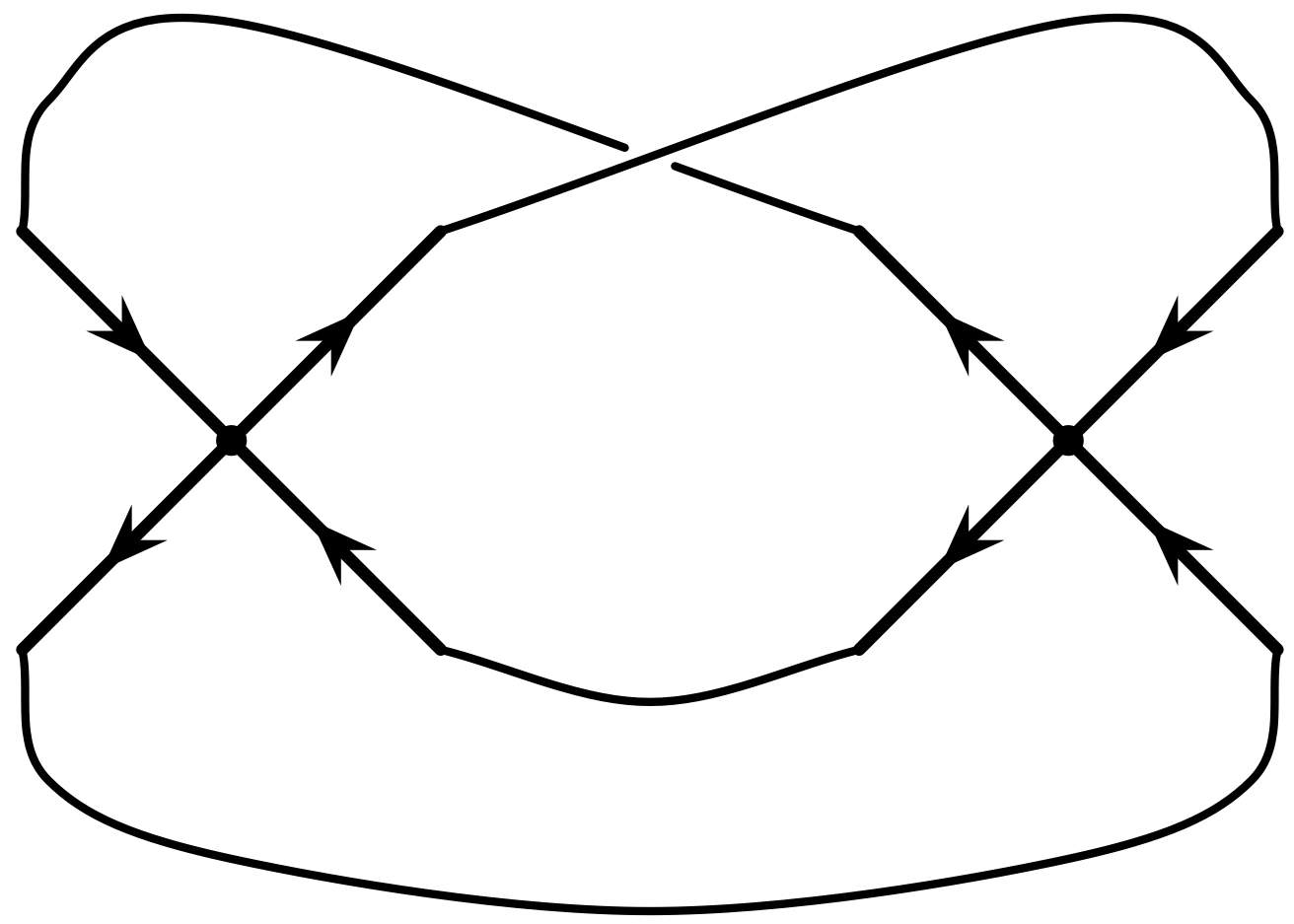}\caption{}\label{fig:hardness_single_saddle}
\end{subfigure}
\begin{subfigure}[b]{0.32\linewidth}
\centering\includegraphics[width=0.75\linewidth]{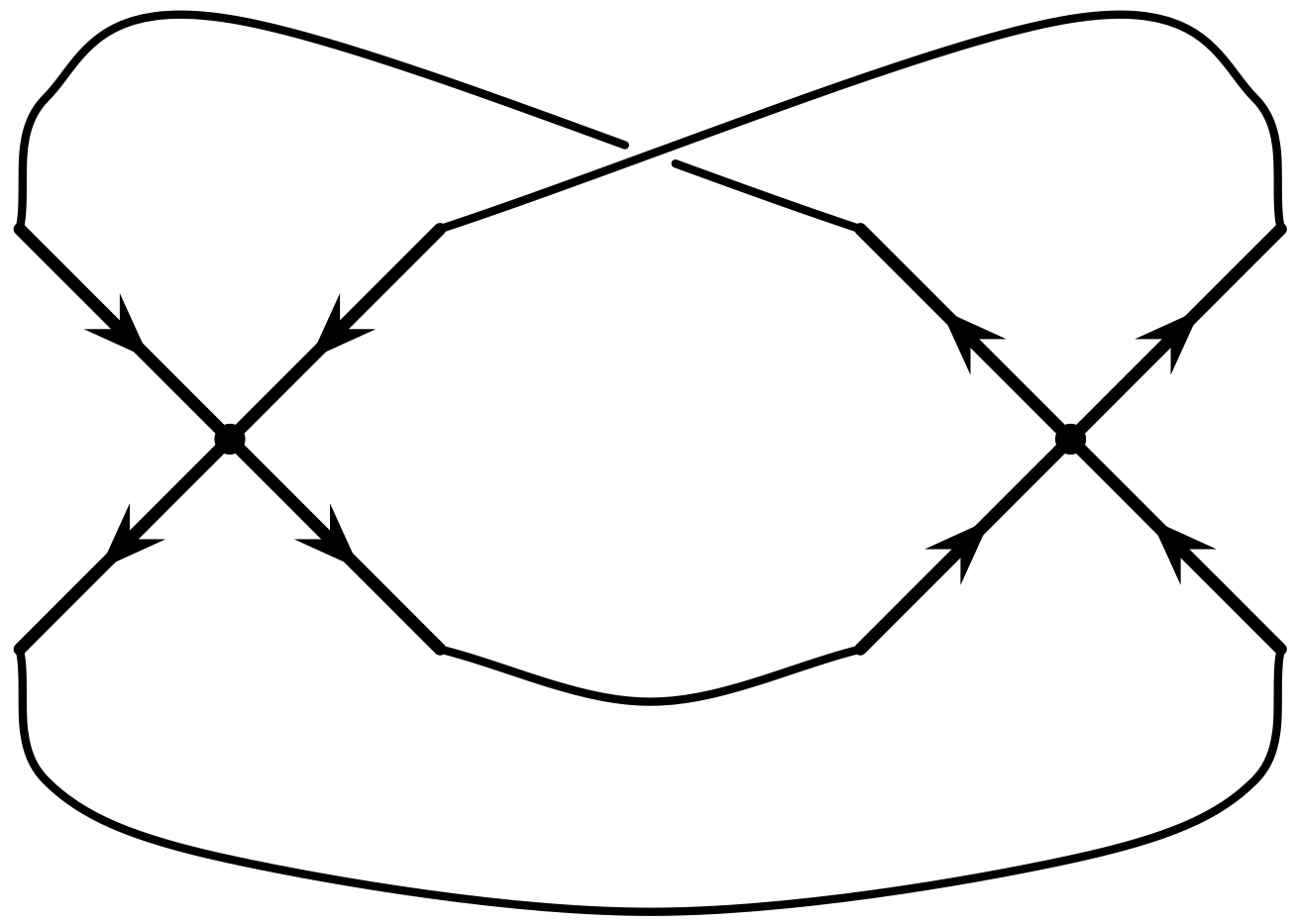}\caption{}\label{fig:hardness_single_nonsaddle}
\end{subfigure}
\caption{A gadget implementing a single edge in independent set.}
\end{figure}

In \figref{fig:hardness_single_undirected} we show how to implement a toy example{\textemdash}a single edge $\{u, v\}${\textemdash}by a gadget of the six-vertex model 
with parameters $(a=1, b=1, c>2)$.
Create two vertices, the left one for $u$ and the right one for $v$, and connect them as is shown in \figref{fig:hardness_single_undirected}.
There are a total of 4 edges. Every 2-in 2-out configuration
on the left vertex uniquely extends to a 2-in 2-out configuration
on the right, and vice versa. Hence there are a total of 6 valid
configurations. When the left vertex has a
 saddle configuration (in-out-in-out, or its reversal) 
which has weight $c$, 
the right must have a non-saddle configuration of weight 1.
\figref{fig:hardness_single_saddle} depicts one such configuration; 
reversing all
arrows gives another one having the same weight. Similarly
if the right has a saddle configuration (or its reversal)
 then  the left  must be
 a non-saddle.
There are two more configurations with two non-saddles
(\figref{fig:hardness_single_nonsaddle} and its reversal).
 This models how two adjacent vertices interact
 in 3-MIS.
We will call the connection pattern 
described in \figref{fig:hardness_single_undirected}
between two sets of 4 dangling edges
 the {\it four-way connection}.
Moreover, when $f$ has
parameters $c > b \ge a = 1$,
 we can label the input wires so that
the 2 saddle configurations of weight $c$  are
paired with the 2 non-saddles
of weight $b$, and the 2 non-saddle/non-saddle pairs
have weight $1$ (by $a=1$).

However, when a vertex in $G$ has more than one neighbors, 
simply duplicating this elementary implementation will not work, because we cannot make sure that the duplicate copies corresponding to the same vertex 
$v$ behave consistently. To handle this difficulty, we design 
a locking gadget (\figref{fig:hardness_arity_three}) for every $v \in V$ 
such that
the property whether $v$ belongs to an independent set in $G$
is consistently reflected in $G'$ in terms of being in
a saddle configuration or not.
This locking mechanism is enforced in the sense of approximation.

\begin{figure}[h!]
\centering
\includegraphics[width=\linewidth]{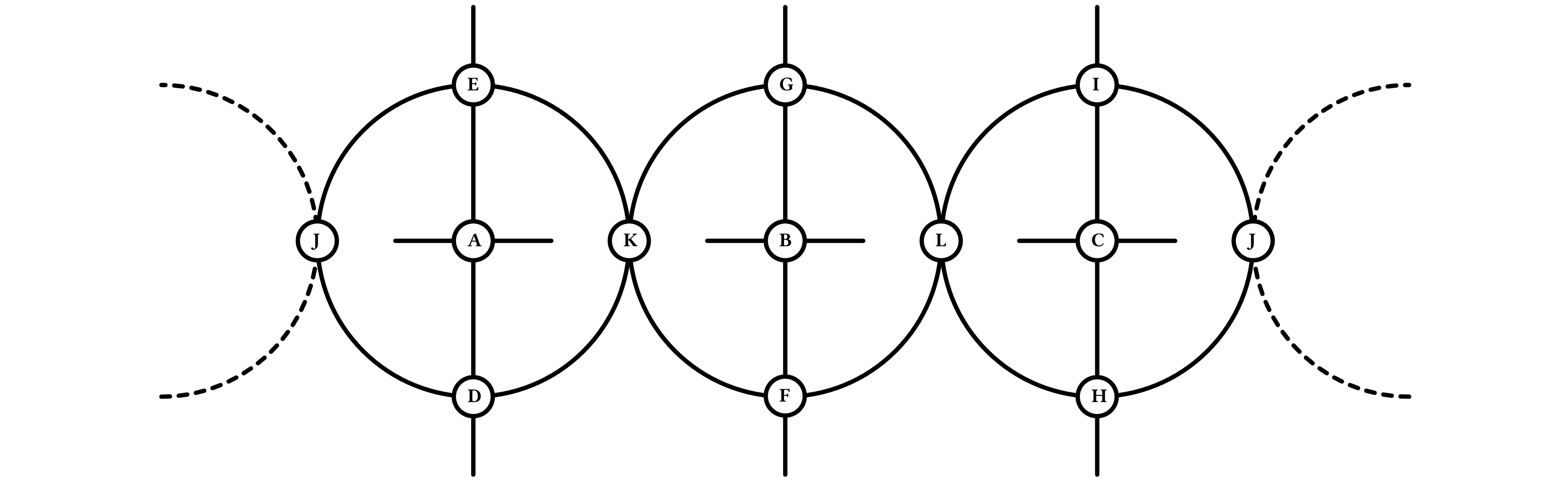}
\caption{A locking 
gadget implementing a degree three vertex and its incident edges.}
\label{fig:hardness_arity_three}
\end{figure}

In \figref{fig:hardness_arity_three}, we identify the leftmost node $J$ with the rightmost node $J${\textemdash}there are three \enquote{circles} in total. 
The nodes will be replaced by a
 RHS gadget in $\Holant\left(\neq_2 | f\right)$.
Each circle has 4 dangling edges.
The \enquote{left circle}
has two dangling edges incident to $A$, one incident to $D$, and one incident to $E$. Similarly for the \enquote{middle circle} and the \enquote{right circle}.
Each edge $\{u, v\}$ in $G$ is modeled by a four-way connection of the 
4 dangling edges between (one circle of the) gadget for $u$ and that for $v$.

The locking mechanism is to realize the following: when the four
dangling edges of one of the 3 circles take a saddle configuration,
(either in-out-in-out, or out-in-out-in), the other two circles must also 
take the identical saddle configuration (in-out-in-out, or out-in-out-in,
respectively); when one circle takes any non-saddle configuration, the other two circles can take independently any non-saddle configurations, with no 
linkage (aside being a non-saddle). 
This is made possible by \textit{chaining},
and the guarantee is enforced by approximate counting.

\begin{figure}[h!]
\centering
\includegraphics[width=0.4\linewidth]{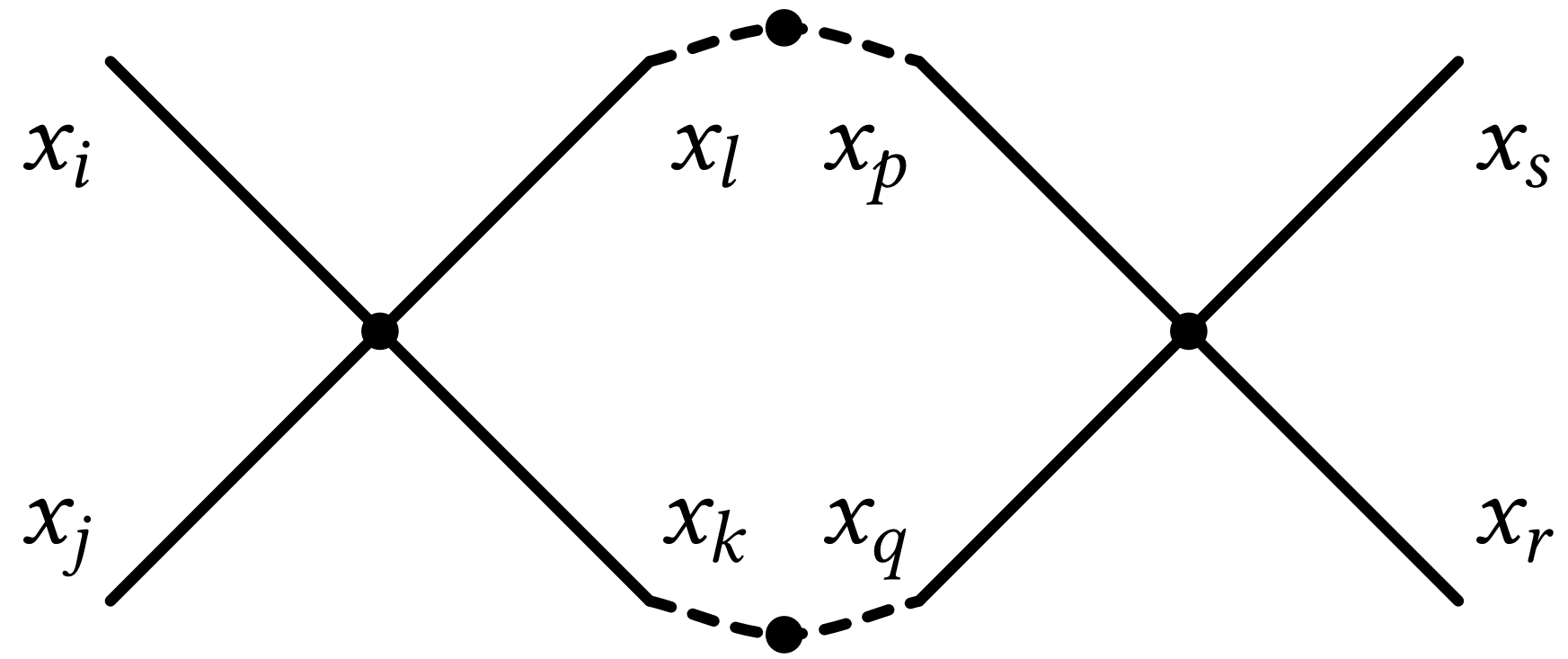}
\caption{Chaining.}
\label{fig:hardness_chain}
\end{figure}

\figref{fig:hardness_chain} depicts a  \textit{2-chain}.
We place the signature $(\neq_2)$ on the two degree 2 vertices connecting
the two degree 4 vertices, each assigned a copy of $f$.
Let $M = M_{x_ix_j,x_lx_k}(f) = M_{x_px_q,x_sx_r}(f) = \left[\begin{smallmatrix} & & & c \\ & a & b & \\ & b & a & \\ c & & &\end{smallmatrix}\right]$.
Then  the signature  $f_2$ of this 2-chain gadget is 
obtained by matrix multiplication $M(f_2) = M_{x_ix_j,x_sx_r}(f_2)
= M N M$,
where $N =  \left[\begin{smallmatrix} & & & 1\\ & & 1 & \\ & 1 & &\\ 1 & & &
\end{smallmatrix}\right]$. Thus
$M(f_2) = \left[\begin{smallmatrix} & & & c_2 \\ & a_2 & b_2 & \\ & b_2 & a_2 & \\ c_2 & & &\end{smallmatrix}\right]
=
\left[\begin{smallmatrix} & & & c^2 \\ & 2ab & a^2+b^2 & \\ & a^2+b^2 & 2ab & \\ c^2 & & &\end{smallmatrix}\right]$,
where
 $c_2 = c^2$,
$a_2 + b_2 = (a+b)^2$. Thus $c_2  > a_2 + b_2$,
and  $\gamma_2 := \frac{c_2}{a_2+b_2} = \left(\frac{c}{a+b}\right)^2 = \gamma^2$.

This can be generalized to a \textit{$k$-chain},  
which connects $k$ vertices with signature $f$ by $k-1$
copies of $N$,  
such that $c_k = c^k, a_k + b_k = (a+b)^k, \gamma_k = \gamma^k$. 
Notice that when $c > a+b$,
the ratio $\frac{c}{a+b}$ can be amplified exponentially in $k$
 in a $k$-chain. Therefore, by a chain of polynomially bounded
size we can ensure the undesirable configurations are negligible{\textemdash}the gadget is locked into the only two complementary configurations which $c$ represents.
It can be verified that $b_k = (\mu^k + \nu^k)/2
 \ge a_k = (\mu^k -\nu^k)/2 \ge 1$, where $\mu = a+b$ and $\nu = b-a$.
We can \enquote{normalize} a $k$-chain by dividing $a_k$,
 so that its parameters are
$\tilde{c}_k = c^k/a_k  > \tilde{b}_k = b_k/a_k \ge \tilde{a}_k =1$.

To reduce the problem 3-MIS to approximating $\Holant\left(\neq_2 | f\right)$, 
let  $\kappa > \lambda \ge 1$ be two constants whose magnitude will later become clear.
For each 3-MIS instance $G = (V, E)$ with $|V| = n$, we construct a graph $G'$ where a gadget in \figref{fig:hardness_arity_three} is created for each $v \in V$, and a four-way connection is made for every $\{u, v\} \in E$,
on the dangling edges 
between two circles corresponding  to $\{u, v\}$ as in
 \figref{fig:hardness_single_undirected}.
For each gadget in \figref{fig:hardness_arity_three},
each of the nodes  $A, B, C$ is replaced by a
\emph{normalized} $\lambda n$-chain to boost the ratio of the saddle configuration over other configurations; each of the nodes
 $D, F, H$  is replaced by a $\kappa n^2$-chain to lock in the configuration \enquote{all arrows pointing up and right} and its reversal; each of  the nodes
 $E, G, I$ is also  replaced by a $\kappa n^2$-chain to lock in the configuration 
\enquote{all arrows pointing down and right} and its reversal
(these configurations at $D, F, H$, and at  $E, G, I$ respectively,
will be called locking configurations); at each of $J, K, L$,
we just put $f$ in which the maximum weight of a configuration over the minimum is a constant $\frac{c}{ \min\{a, b\}} = c$.
Note that the signature in \figref{fig:hardness_chain} 
has the dominating entry at 0011 and 1100. Since our graph $G'$ does not
need to be planar, we can reorder the 4 external edges arbitrarily.
In particular, for  $A, B, C$ the dominating entry $\tilde{c}_{\lambda n}$ is
in the saddle 0101 and 1010 positions, 
as depicted in \figref{fig:hardness_arity_three}.
Similarly the  4 external edges of $D, F, H$  and $E, G, I$ are
also properly reordered, from the order given in 
 \figref{fig:hardness_chain}, as an $\lambda n$-chain to achieve
the proper locking configurations.
 
Next we argue that the maximum size $s$ of independent sets
 in $G$ can be recovered from an approximate solution to
$\Holant\left(G'; \neq_2 | f\right)$.

Given an independent set $S \subset V$ of size $s$, we show there is a 
valid configuration (at the granularity of nodes
and edges shown in \figref{fig:hardness_arity_three}) 
of weight $\ge c^{6\kappa n^3} \left(\tilde{c}_{\lambda n} \tilde{b}_{\lambda n}\right)^{3s}$. 
For any vertex  $v \in S$ we set the following configuration for its locking
gadget: 
 set each of 3 nodes $A, B, C$ to the same 
saddle configuration  in-out-in-out cyclically starting from the upper 
edge{\textemdash}each has weight $\tilde{c}_{\lambda n}$; 
set each of 3 nodes  $D, F, H$
 to the same out-out-in-in locking configuration (clockwise) 
cyclically starting from the upper edge{\textemdash}each has weight $c^{\kappa n^2}$; 
set  each of 3 nodes  $E, G, I$ to the same 
in-out-out-in locking configuration (clockwise)     
cyclically starting from the upper edge{\textemdash}each also has weight $c^{\kappa n^2}$;
set  each of 3 nodes  $J, K, L$
to the same configuration \enquote{two in from the left and two out to the right},
 which has a non-zero weight $\ge 1$.
For any vertex $v \not \in S$
 we set the following configuration for its locking
gadget:
All $D, F, H, E, G, I$ will be in some  locking configurations.
Consider any of the 3 circles in the gadget,
 for example the circle formed by $A, D, E, J, K$. The node $A$ is
involved in a  four-way connection to another  circle belonging to
a gadget for some vertex $u$. If $u \in S$, the assigned configuration
just defined at $u$ forces a non-saddle configuration here;
more specifically the horizontal two dangling edges at $A$ must 
either both point right or both point left, and
the upper edge $\hat{e}$ of $E$ and lower edge $\underline{e}$ of $D$ must 
either both point up or both point down. 
Regardless of which of the two assignments for  $\hat{e}$  and  $\underline{e}$
we can assign a locking configuration for $E$ and $D$
so that the upper and lower edges of $A$ are
either both point up or both point down. 
%
Note that in either case, the left two edges of $K$ are one-in-one-out;
similarly
the right  two edges of $J$ are also one-in-one-out
(this allows \enquote{freedom} between the 3 circles where each of $J$, $K$, $L$
can take a nonzero weight $\ge 1$).  
Continuing at the circle $A, D, E, J, K$,
 if $u \not \in S$, then we will pick an arbitrary
non-saddle to non-saddle configuration in the 4-way connection for $\{u, v\}$.
These can all be extended to a valid configuration at $A, D, E$
such that the configuration at $A$ is non-saddle having weight $\ge 1$,
the configurations at $D$ and $E$ are locking,
and the right two  edges of $J$  and the  left two edges of $K$ are
both one-in-one-out.
The  weight  at  $D$ and $E$ are still $c^{\kappa n^2}$.
Because $J$ and $K$ each has one-in-one-out from within the side of the 
circle, the 3 circles can be assigned independently from each other.  This
allows us to handle the situation where, for the same $v \not \in S$,
 some edge $\{v, u\}$
connects to $u \in S$ and some edge $\{v, u'\}$ connects to $u' \not \in S$.

We have defined a valid configuration, and it has weight  
$\ge \prod_{v \in V}c^{6\kappa n^2}
\prod_{v \in S} \left(\tilde{c}_{\lambda n} \tilde{b}_{\lambda n}\right)^{3}
= c^{6\kappa n^3} \left(\tilde{c}_{\lambda n} \tilde{b}_{\lambda n}\right)^{3s}$,
 where $6$ comes from the 6 locking nodes
$D, E, F, G, H, I$ in each locking gadget.
(Omitted factors are all $\ge 1$.)

Next we show the weighted sum of all configurations 
is smaller than $c^{6\kappa n^3} 
\left(\tilde{c}_{\lambda n} \tilde{b}_{\lambda n}\right)^{3(s+1)}$.
First we bound  $W_{\rm lock}$, the sum of weights for configurations
where all nodes labeled $D, E, F, G, H, I$ are locked.
Consider any circle such as the one labeled 
 $A, D, E, J, K$ for any $v \in V$. It is involved in  a four-way
connection with another such circle for a vertex $u \in V$, say
$A', D', E', J', K'$,  where
$\{u,v\} \in E$.
That $D$ is locked forces that
the  upper and lower edges of $D$ are to be consistently oriented,
i.e., both up or both down. Similarly, consistency holds
at $E$, $D'$ and $E'$. Thus the four-way
connection forces that there can be at most one of
$A$ and $A'$ is in a saddle configuration. 
Furthermore, if $A$ is in a particular saddle configuration,
say in-out-in-out starting from the upper edge,
both  upper and lower edges of $D$ must point up, and
both  upper and lower edges of $E$ must point down,
and then both right edges of $D$ and $E$ must point right,
causing the left two edges of $K$ point in, and thus the right
 two edges of $K$ point out. This forces both $F$ and $G$
to take \emph{exactly} the same locked configurations
of $D$ and $E$ respectively, whch forces $B$ to be in
\emph{exactly} the same  saddle configuration as $A$.
Similarly so is $C$. We conclude that when all
nodes labeled $D, E, F, G, H, I$ are locked, for any $v \in V$,
if any of its $A, B, C$ is in a saddle configuration, then
all 3 are in exactly the same saddle configuration,
and none of $A', B', C'$ for $u \in V$ is a saddle,
if $\{u,v\} \in E$. In particular, there can be at most $3s$ many
saddles among $A, B, C$'s in $G'$.
If $0 \le i \le s$ is the number of $\{A, B, C\}$'s being in saddle,
their weight is $\left(\tilde{c}_{\lambda n}\right)^{3i}$, and
their corresponding non-saddles  in respective four-way
connections must take weight 
 $\left(\tilde{b}_{\lambda n}\right)^{3i}$.
Those  $(3n-6i)/2$ pairwise four-way
connections between two non-saddles have weight $\tilde{a}_{\lambda n} =1$.
Note that, if any of those non-saddles were to take weight
$\tilde{b}_{\lambda n}$, then the corresponding paired node
in its four-way
connection must be in saddle, a contradiction.

It follows that
\[W_{\rm lock} \le 2^{6n} \left(c^{\kappa n^2}\right)^{6n} 
 \sum_{i=0}^s\binom{n}{i}
\left(\tilde{c}_{\lambda n} \tilde{b}_{\lambda n}\right)^{3i} c^{3n} 
\le 2^{7n} c^{6\kappa n^3 + 3n} 
\left(\tilde{c}_{\lambda n} \tilde{b}_{\lambda n}\right)^{3s},\]
where each locked node $D, E, F, G, H, I$ has 2 possible locking
configurations each with weight  $c^{\kappa n^2}$,
and given a particular assignment of  $6n$ locking configurations, 
there can be at most $s$ batches of
$A, B, C$'s in saddle configurations (same for each batch and determined by 
the locks)  with weight $\tilde{c}_{\lambda n}$.
Hence $W_{\rm lock} < \frac{1}{2}c^{6\kappa n^3} 
\left(\tilde{c}_{\lambda n} \tilde{b}_{\lambda n}\right)^{3(s+1)}$,
when 
$\lambda \ge 1 $ is large.


It remains to upper-bound the weighted sum of configurations where there is at least one gadget with some lock broken. This quantity is bounded by
\begin{align*}
& \sum_{i = 0}^{6n - 1}\binom{6n}{i}c^{i\kappa n^2}(a+b)^{(6n-i)\kappa n^2}
6^{6n}
\left[2\left(\tilde{c}_{\lambda n} + \tilde{b}_{\lambda n} + 1\right)\right]^{3n}[2(a+b+c)]^{3n} \\
\le \; & 2^{\Theta(n)} (a+b)^{6 \kappa n^3}
\left( \frac{c}{a+b} \right)^{(6n-1)\kappa n^2}
 \left[\tilde{c}_{\lambda n} + \tilde{b}_{\lambda n} + 1\right]^{3n}[a+b+c]^{3n}\\
\le \; & 2^{\Theta(n)} [\Theta(1)]^{\lambda n^2} c^{6 \kappa n^3}
\frac{1}{ c^{\kappa n^2}} (a+b)^{\kappa n^2}\\
\le \; & 2^{\Theta(n)} [\Theta(1)]^{\lambda n^2} 
c^{6\kappa n^3} \frac{1}{\gamma^{\kappa n^2}},
\end{align*}
which is  $< \frac{1}{2}c^{6\kappa n^3}$
when $\kappa \gg \lambda \ge 1 $ is large.
\end{proof}

%% file: open.tex
\documentclass[paper]{subfiles}

The main open problem on the approximate complexity of the six-vertex model
is in the white region.
The finer classification of the approximate complexity for the planar case
is also open. Approximating $T(G; 3,3)$ is \#BIS-hard for general graphs~\cite{Goldberg:2012:APF:2371656.2371660}. On planar graphs, $T(G; 3,3)$ is equivalent to the six-vertex model at $(1, 1, 2)$ where the approximation complexity for planar graphs is unknown.

%% file: appendix.tex
\documentclass[paper]{subfiles}

The \textit{near-assignments Markov chain} for \textit{windable functions} is proposed by Colin McQuillan in \cite{DBLP:journals/corr/abs-1301-2880} and further analyzed in \cite{Huang:2016:CPM:2884435.2884473}.
In the following we present the definition of windable functions given in the latter and show that signatures in $\mathcal{F}_{\le^2}$ are all windable. 
This
leads to an alternative derivation of a rapidly mixing Markov chain,
and based on this chain
the techniques introduced in \secref{sec:fpras} can be adapted to
give a FPRAS for $\Holant\left(\neq_2 | \mathcal{F}_{\le^2}\right)$.
\begin{defn*}\label{windable}
For any finite set $J$ and any configuration $x \in \{0,1\}^J$, define $\mathcal{M}_x$ to be the set of partitions of $\{i \ | \ x_i = 1\}$ into pairs and at most one singleton. A signature $f: \{0,1\}^J \rightarrow \mathbb{R^+}$ is \textit{windable} if there exist values $B(x,y,M) \ge 0$ for any two
distinct $x, y \in \{0,1\}^J$ and all $M \in \mathcal{M}_{x \oplus y}$, such 
that:
\setlist[enumerate]{itemsep=-1mm}
\begin{enumerate}[(I)]
\item
$f(x)f(y) = \sum_{M \in \mathcal{M}_{x \oplus y}}B(x, y, M)$ for all
distinct  $x, y \in \{0, 1\}^J$, and
\item
$B(x, y, M) = B(x \oplus S, y \oplus S, M)$ for all distinct
 $x, y \in \{0, 1\}^J$ and all $S \in M \in \mathcal{M}_{x \oplus y}$.
\end{enumerate}
Here $x \oplus S$ denotes the vector obtained by changing $x_i$ to $1 - x_i$ for the one or two elements $i$ in $S$.
\end{defn*}

The signature $(\neq_2)$ is shown to be windable in \cite{DBLP:journals/corr/abs-1301-2880}. We show that signatures in $\mathcal{F}_{\le^2}$ are all windable.
\begin{lem*}
For any nonnegative real numbers $a$, $b$, and $c$, the function $f$ with signature matrix $M(f) = \left[\begin{smallmatrix} & & & a \\ & b & c & \\ & c & b & \\ a & & &\end{smallmatrix}\right]$ is windable if and only if $a^2 \le b^2 + c^2$, $b^2 \le a^2 + c^2$, and $c^2 \le a^2 + b^2$.
\end{lem*}
\begin{proof}
According to the definition, we need to verify when there exist values $B(x, y, M) \ge 0$ for all distinct $x, y \in \{0,1\}^4$ and all $M \in \mathcal{M}_{x \oplus y}$ that satisfy the two conditions. 
Let  $\operatorname{Hw}(x)$ denote the Hamming weight of a 0-1 vector $x$. We make the following observations:
\setlist[enumerate]{itemsep=-1mm}
\begin{enumerate}[(1)]
\item
Since $f$ only takes nonzero values on inputs of Hamming weight ($\operatorname{Hw}$) 2,
by \textit{condition I} we must have $B(x, y, M) = 0$ for 
all $M \in \mathcal{M}_{x \oplus y}$,
if $\operatorname{Hw}(x) \neq 2$ or $\operatorname{Hw}(y) \neq 2$.
\item
For any distinct $x, y$ with $\operatorname{Hw}(x) = \operatorname{Hw}(y) = 2$, if $y \neq \bar{x}$ then $|\mathcal{M}_{x \oplus y}| = 1$. Denote this unique partition by $M$, by \textit{condition II} we have $B(x, y, M) = B(y, x, M)$.
\item
For any  $x, y$ with $\operatorname{Hw}(x) = \operatorname{Hw}(y) = 2$, if $y = \bar{x}$ then
\[\mathcal{M}_{x \oplus y} = \{\{\{1,2\}, \{3,4\}\}, \{\{1,3\}, \{2,4\}\}, \{\{1,4\}, \{2,3\}\}\}\]
and we still have $B(x, y, M) = B(y, x, M)$ for any $M \in \mathcal{M}_{x \oplus y}$.
\item
B(0011, 1100, \{\{1,2\}, \{3,4\}\}) = B(0110, 1001, \{\{1,4\}, \{2,3\}\}) = B(0101, 1010, \{\{1,3\}, \{2,4\}\}) = 0; otherwise, by \textit{condition II} 
and \textit{condition I} we 
would have $f(0000)f(1111) \neq 0$, a contradiction.
\item
By \textit{condition II},
\setlist[itemize]{itemsep=-1mm}
\begin{itemize}
\item
\(B(0011, 1100, \{\{1,3\}, \{2,4\}\}) = B(0110, 1001, \{\{1,3\}, \{2,4\}\})\);
\item
\(B(0011, 1100, \{\{1,4\}, \{2,3\}\}) = B(0101, 1010, \{\{1,4\}, \{2,3\}\})\);
\item
\(B(0110, 1001, \{\{1,2\}, \{3,4\}\}) = B(0101, 1010, \{\{1,2\}, \{3,4\}\})\).
\end{itemize}
Denote them by $X$, $Y$, and $Z$, respectively.
\end{enumerate}
Combining these observations, we are left with 15 variables and 15 equations:
\[ 
\left\{
\begin{array}{ll}
B(0011, 0110, \{\{2,4\}\}) = f(0011)f(0110) = ab, & B(0110, 0101, \{\{3,4\}\}) = f(0110)f(0101) = bc,\\
B(0011, 1001, \{\{1,3\}\}) = f(0011)f(1001) = ab, & B(0110, 1010, \{\{1,2\}\}) = f(0110)f(1010) = bc,\\
B(0011, 0101, \{\{2,3\}\}) = f(0011)f(0101) = ac, & B(1001, 0101, \{\{1,2\}\}) = f(1001)f(0101) = bc,\\
B(0011, 1010, \{\{1,4\}\}) = f(0011)f(1010) = ac, & B(1001, 1010, \{\{3,4\}\}) = f(1001)f(1010) = bc,\\
B(1100, 0110, \{\{1,3\}\}) = f(1100)f(0110) = ab, & X + Y = f(0011)f(1100) = a^2,\\
B(1100, 1001, \{\{2,4\}\}) = f(1100)f(1001) = ab, & X + Z = f(0110)f(1001) = b^2,\\
B(1100, 0101, \{\{1,4\}\}) = f(1100)f(0101) = ac, & Y + Z = f(0101)f(1010) = c^2,\\
B(1100, 1010, \{\{2,3\}\}) = f(1100)f(1010) = ac.\\
\end{array}
\right.
\]
(Here the $B$'s and $X, Y, Z$ are variables.)
The system of equations has nonnegative solutions if and only if the set of the last three equations has. This holds if and only if  $a^2 \le b^2 + c^2$, $b^2 \le a^2 + c^2$, and $c^2 \le a^2 + b^2$.
\end{proof}